\documentclass[sigconf]{acmart}

\usepackage[english]{babel}
\usepackage{blindtext}
\usepackage{graphicx}
\usepackage{mathrsfs}
\usepackage{amsmath}
\usepackage{amsfonts}
\usepackage{xcolor}

\usepackage{multirow}
\usepackage{comment}
\usepackage{flushend}
\usepackage{epstopdf}
\usepackage{booktabs}
\usepackage{url}
\usepackage{cleveref}

\usepackage{float}
\usepackage{subfigure}
\usepackage{caption}
\usepackage{natbib}


\usepackage{chngcntr}

\usepackage{algorithm}
\usepackage{algorithmic}

\renewcommand\footnotetextcopyrightpermission[1]{} 
\setcopyright{none}

\acmDOI{}

\acmISBN{}

\acmPrice{}

\AtBeginDocument{%
  \providecommand\BibTeX{{%
    \normalfont B\kern-0.5em{\scshape i\kern-0.25em b}\kern-0.8em\TeX}}}

\setcopyright{acmcopyright}
\copyrightyear{2018}
\acmYear{2018}
\acmDOI{10.1145/1122445.1122456}

\acmConference[]{}{}{}
\acmBooktitle{}
\acmPrice{}
\acmISBN{}

\begin{document}
\newenvironment{shrinkeq}[1]
{ \bgroup
\addtolength\abovedisplayshortskip{#1}
\addtolength\abovedisplayskip{#1}
\addtolength\belowdisplayshortskip{#1}
\addtolength\belowdisplayskip{#1}}
{\egroup\ignorespacesafterend}
\title{{\it ANT}: Learning Accurate Network Throughput for Better Adaptive Video Streaming}

\author{Jiaoyang Yin, Yiling Xu, Hao Chen, Yunfei Zhang, Steve Appleby, Zhan Ma}

\begin{abstract}
  Adaptive Bit Rate (ABR) decision plays a crucial role for ensuring satisfactory Quality of Experience (QoE) in video streaming applications, in which past network statistics are mainly leveraged for future network bandwidth prediction. However, most algorithms, either rules-based or learning-driven approaches, feed throughput traces or classified traces based on traditional statistics (i.e., mean/standard deviation) to drive ABR decision, leading to compromised performances in specific scenarios. Given the diverse network connections (e.g., WiFi, cellular and wired link) from time to time, this paper thus proposes to learn the {\it ANT} (a.k.a., Accurate Network Throughput) model to characterize the full spectrum of network throughput dynamics in the past for deriving the proper network condition associated with a specific cluster of network throughput segments (NTS). Each cluster of NTS is then used to generate a dedicated ABR  model, by which we wish to better capture the network dynamics for diverse connections.  We have integrated the {\it ANT} model with existing reinforcement learning (RL)-based ABR decision engine, where different ABR models are applied to respond to the accurate network sensing for better rate decision. Extensive experiment results show that our approach can significantly improve the user QoE by 65.5\% and 31.3\% respectively, compared with the state-of-the-art Pensive and Oboe, across a wide range of network scenarios.
\vspace{-2mm}
\end{abstract}


\maketitle

\section{Introduction}
Recent years have witnessed an exponential increase of HTTP-based video streaming traffic volume~\cite{forecast2019cisco}. To assure the high-quality service provisioning, adaptive bit rate (ABR) policy is  devised to select appropriate video chunks for combating the network dynamics (e.g., packet loss, congestion, throughput fluctuation, etc) from time to time  to achieve satisfactory QoE for heterogeneous connections.

Prevalent ABR algorithms, i.e., either rules-based or recently emerged learning-based approaches~\cite{liu2011rate, jiang2012improving, sun2016cs2p, miller2016qoe, kurdoglu2016real, zou2015can,huang2015buffer, huang2012confused, huang2013downton, miller2012adaptation, spiteri2016bola, beben2016abma,tian2012towards, zhou2013buffer, li2014probe, wang2016squad, mansy2013sabre, yin2014toward,  yin2015control, de2013elastic, mao2017neural, akhtar2018oboe, huang2020stick, hong2019continuous, jiang2019hd3, huang2019comyco, peng2019hybrid, huang2018qarc, elgabli2020fastscan,mao2017neural}, typically run on client-side to perform the video bit rate decision through client-side playback buffer occupancy observation, and/or network throughput (bandwidth) estimation. Since existing ABR methods
typically model the average of different network throughput traces for bit rate prediction using fixed control mechanism or very limited neural models, they mostly overlook the significant variations of network connections from time to time and from device to device. For example, Pensive~\cite{mao2017neural}, a pioneering learning-based ABR method, characterized the averaged network behavior from a number of different network traces to drive a single neural bit rate decision model. Recently, Oboe~\cite{akhtar2018oboe} suggested two ABR decision models for rate adaptation to cover two different bit rate ranges respectively, where each ABR decision model was trained using corresponding traces belonging to the same bit rate range.

Simply averaging network throughput is not sufficient to represent its variations for any connection. Thus we propose the {\it ANT} model to learn accurate network throughput dynamics for better adaptive video streaming.  Towards this, we first attempt to classify the network into multiple (e.g., five) clusters by applying K-means to analyze the characteristics of network traces, by which each cluster of NTS, a.k.a., network condition, is assumed to represent a typical network connection having its discriminative behavior.
Then, we map past NTS to a specific network condition by quantitatively measuring its feature vector, where we leverage the deep neural network (DNN) to learn temporal dynamics instead of simply using the mean and/or standard deviation (std) of NTS~\cite{akhtar2018oboe}. In the meantime,  we further use clustered NTS (as of different network conditions) from network traces to train corresponding neural models for ABR decision, with which we can leverage multiple neural ABR models to accurately represent diverse network connections. In this work, we train well-known and emerging RL-based ABR decision engine to better select bit rate for improved QoE~\cite{mao2017neural,akhtar2018oboe,ChenTgaming}.

{\bf Contributions.}
1) We suggest the ANT model to classify incoming NTS into one of five discriminative network conditions to drive corresponding neural ABR decision model, by which we can better and accurately represent temporal network throughput dynamics that are often overlooked in literature.

2) Instead of simply applying the mean and/or std of incoming NTS to recognize different network conditions, we rely on DNNs to accurately characterize the feature representation of NTS dynamics over time, by which we could better differentiate typical scenarios of network connections.

3) We demonstrate our ANT model using various public accessible network traces and alternative Tencent traces obtained from large-scale Tencent video hosting platform, having noticeable QoE improvements by 65.5\% and 31.3\% respectively against the state-of-the-art Pensive~\cite{mao2017neural} and Oboe~\cite{akhtar2018oboe}.

\vspace{-2mm}

\section{Background and Motivation}
  Various ABR algorithms
  are adopted in popular video streaming applications. They mostly collect network throughput~\cite{liu2011rate, jiang2012improving, sun2016cs2p, miller2016qoe} and receiver states (e.g., buffer status, packet loss, etc)~\cite{huang2015buffer, huang2013downton, miller2012adaptation, spiteri2016bola} in the past for future rate adaptation.

  In reality, it is hard for conventional heuristic methods to sufficiently cover network variations.
  Thus, recent years have witnessed the growth of learning-based approaches, in which well-known RL-based neural engine is often deployed to learn a single model to represent diverse network behaviors from a great number of public network traces~\cite{mao2017neural,ChenTgaming}. In this way, learned neural model has to compromise among all possible network conditions, typically leading to  acceptable but not the best performance.

\begin{figure}[htbp]
\centering
\includegraphics[width=3.3in, height=2.8in]{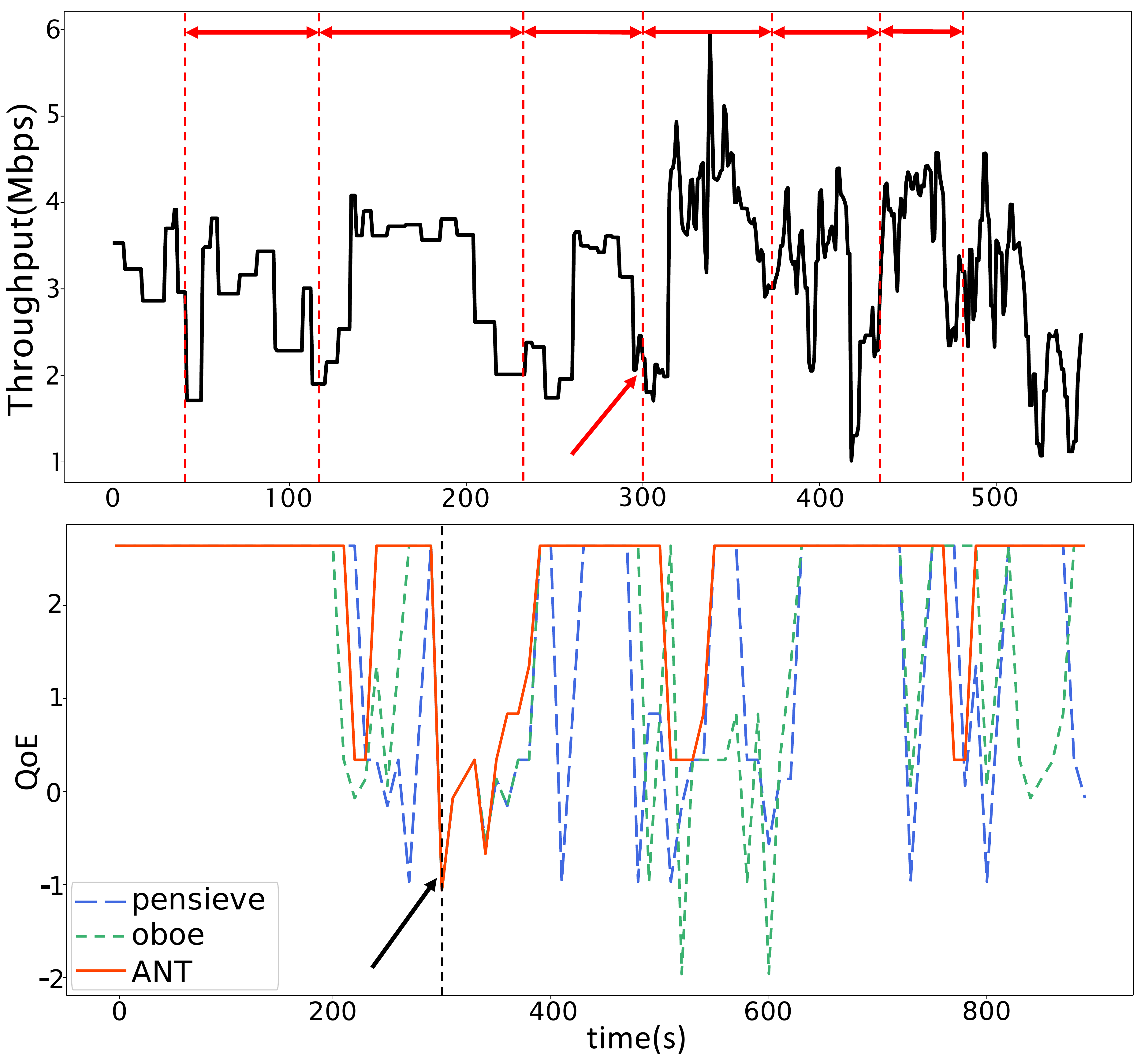}
\caption{Necessity for network throughput learning.}
\label{figure:1}
\vspace{-2mm}
\end{figure}

In order to deal with various network conditions during transmission, ~\cite{akhtar2018oboe} proposed Oboe which can auto-tune video ABR algorithm to network conditions. Oboe detects changes in network states/conditions using Bayesian change point detection algorithms based on average and std of throughput, and adjusts ABR parameters according to detection results. Benefiting from the auto-tuning mechanism, ABR algorithms can adapt to complex network conditions and obtain optimal performance in QoE metric and underlying metrics. However, Oboe detects and recognizes network condition changing based on average and std of throughput essentially, which may not be enough to accurately represent dynamic and complex network conditions. And in this way, algorithm will not auto-select appropriate ABR parameters. As shown in Fig.~\ref{figure:1}, throughput trends with similar average and std values appear during time slots between red dashed lines. When adopting fixed ABR model or distinguishing throughput trends according to average or std value, subtle differences in throughput trends may not be recognized and appropriate ABR model can not be chosen in time, and hence, incurring QoE fluctuation. On the contrary, if network throughput is analyzed and learned accurately, as proposed {\it ANT} shows, appropriate ABR model is chosen in time and QoE fluctuation will be prevented to some extent. To this end, it is necessary to establish a mechanism to learn accurate network throughput and recognize comprehensive changes in throughput trends.

\section{Learning {\it ANT} Model and Its Application to ABR Decision}
\begin{figure}[t]
\centering
\includegraphics[width=3.4in]{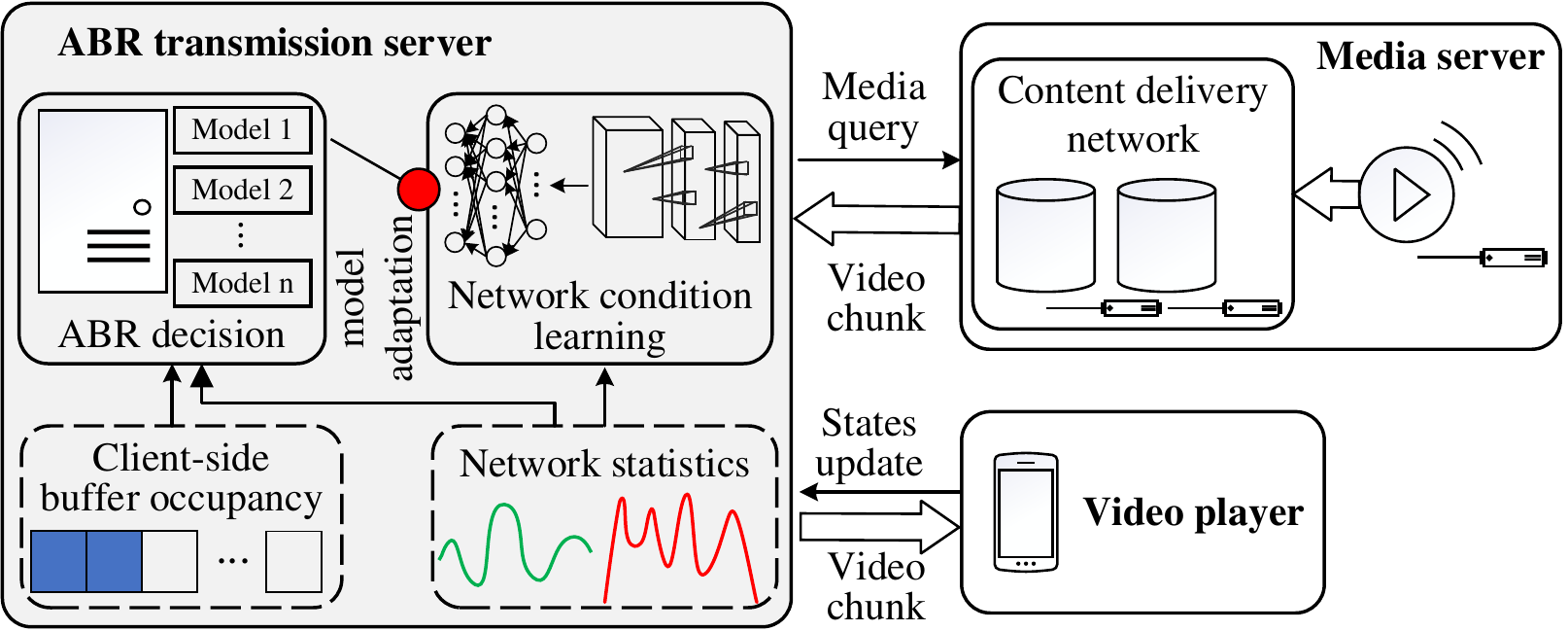}
\caption{Overall architecture of {\it ANT}-powered adaptive video streaming.}
\label{figure:2}
\vspace{-2mm}
\end{figure}
\subsection{Architecture Overview}
The overall architecture of {\it ANT}-powered adaptive video streaming is shown in Fig.~\ref{figure:2}. It consists of a network condition learning module and a condition-wised ABR decision module. The input of network condition learning module is the historical throughput data obtained at client-side. Then a Convolutional Neural Network (CNN) is utilized to extract features of historical network data automatically and output recognition results of future network conditions after analyzing. ABR decision module adopts various states, such as historical bit rate, current buffer size, past chunk throughput, past downloaded time, future chunk size, as inputs. Then these states are fed into one of multiple neural network models trained under various network conditions, based on the recognition result of network condition learning module, to make ABR decision. The decisions made by the neural network will affect the video streaming environment, and in reverse, the environment will feedback changed states into the ABR decision module. The overall procedure for proposed architecture can be formulated as follows:
\begin{align}\label{1}
condition = {f_{1D - CNN}}(throughput{_{historical}})
\vspace{-2mm}
\end{align}
\begin{align}\label{2}
action = {f_{ABR}}(stat{e_{network}},stat{e_{player}},condition)
\vspace{-2mm}
\end{align}

\subsection{Network Condition Learning Module}\label{ssec:condition_learning}
 The network condition learning module we construct is based on CNN, which has the ability to extract comprehensive features of input data automatically. We propose a power CNN framework to analyze historical throughput data and learn current network conditions for assisting future model selection. Historical network throughput data is adopted as input of this module to perform both training and testing phases.

 \begin{figure}[b]
\centering
\includegraphics[width=3.5in]{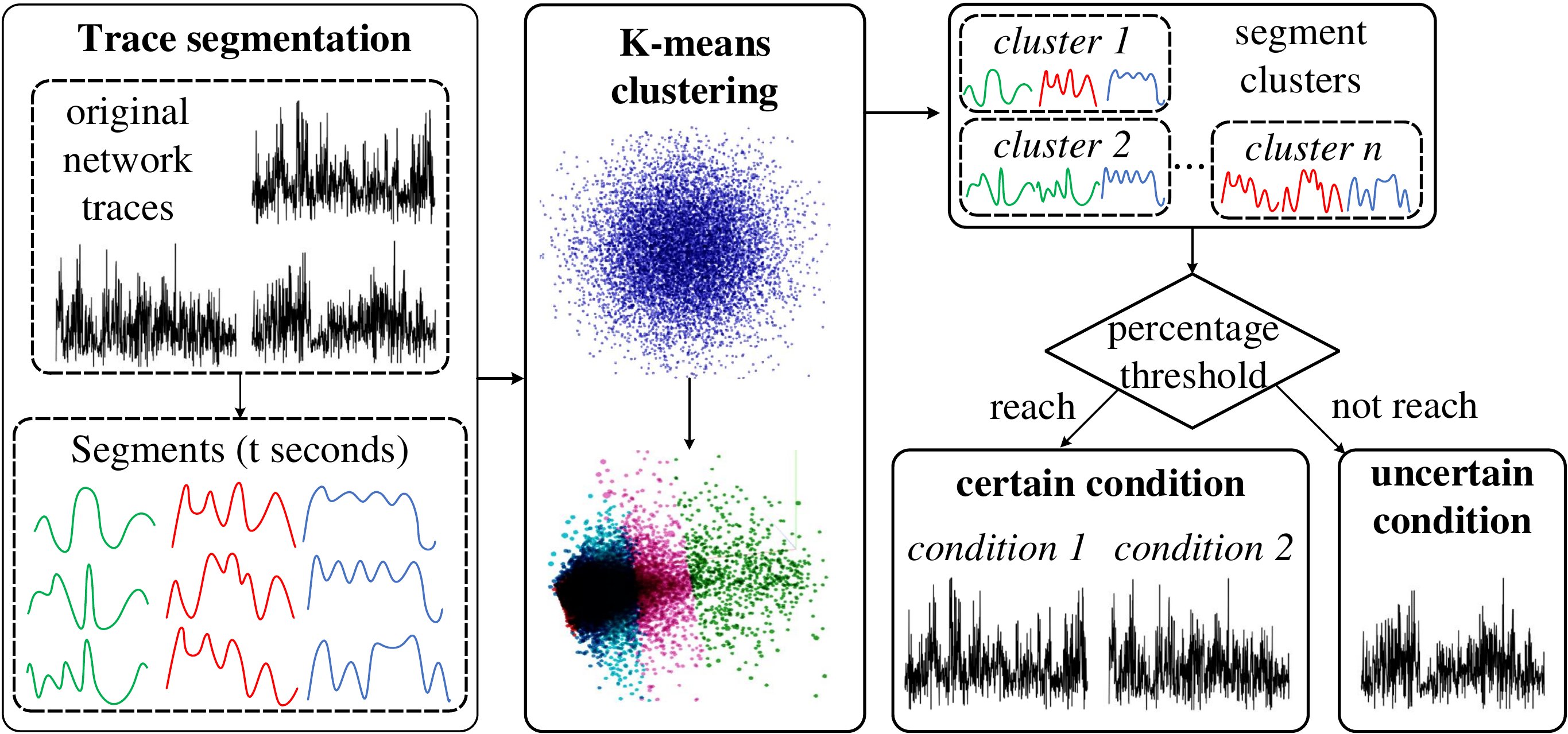}
\caption{Illustration of the trace aggregation mechanism.}
\label{figure:3}
\vspace{-2mm}
\end{figure}

 \begin{figure*}[ht]
\centering
\includegraphics[width=6in]{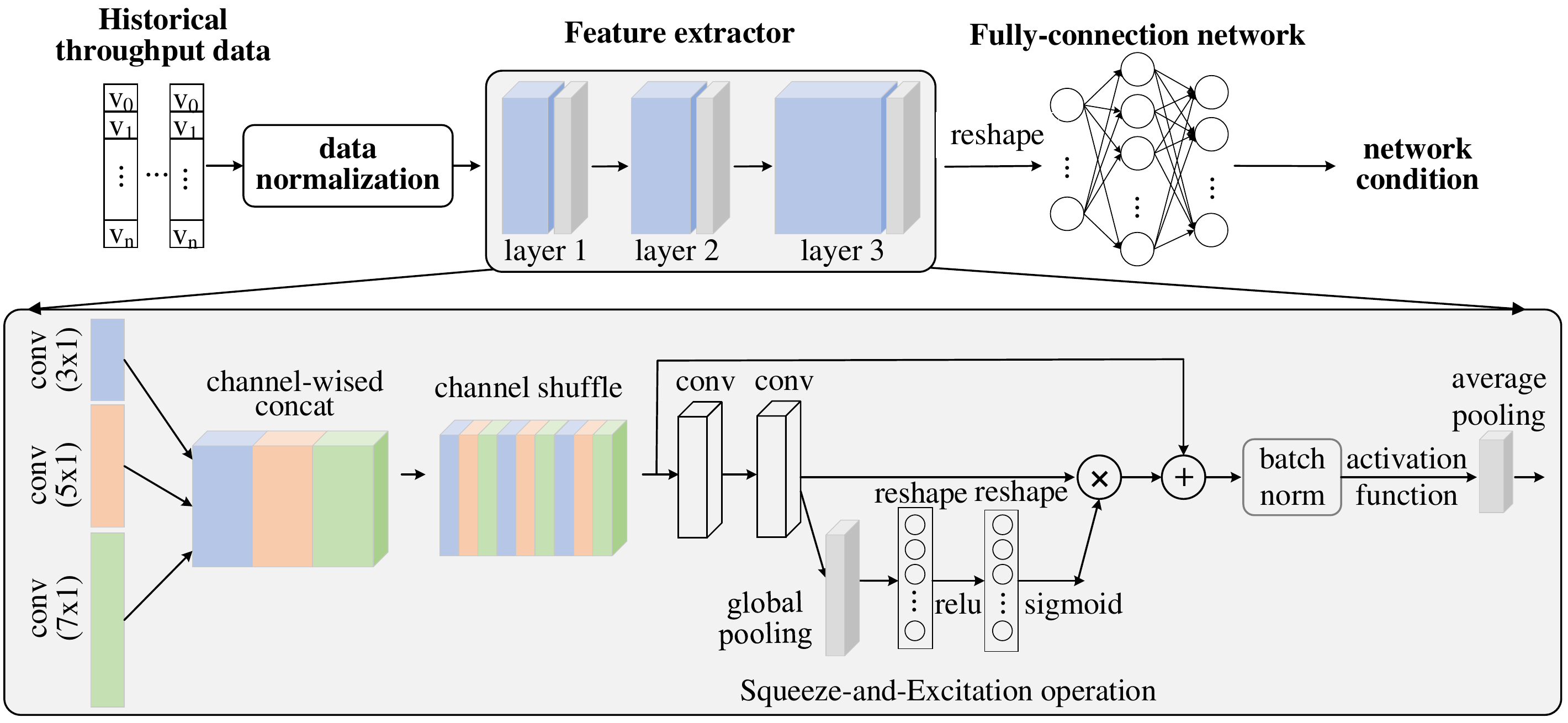}
\caption{Framework of network condition learning module.}
\label{figure:4}
\vspace{-2mm}
\end{figure*}

 Due to lacking related labels in current network traces to indicate real network conditions, it will encounter difficulty when training and validating this CNN. To tackle this issue, we propose a trace aggregation mechanism to distinguish network traces based on the data feature in temporal dimension, as shown in Fig.~\ref{figure:3}. Original network traces are firstly split into several segments which contain temporal information of $t$ seconds. Then K-means~\cite{1966Some}, a typical clustering algorithm, is adopted to achieve segment aggregation according to the similarity of temporal features in $t$ seconds' segment. After K-means operation, throughput segments from various network traces are aggregated into $k$ clusters, and segments with similar network condition are classified into same cluster. Eventually, the whole network trace can be labeled by the most frequent type of network conditions for all component trace segments. It should be noted that if the percentage (frequency) does not reach a predefined threshold $h$, ``{\it uncertain}'' will be marked for this trace. With the network condition labels tagged for the throughput data, these network traces can be fed into the CNN to conduct training and testing.

 With the indication of condition labels, the neural network is trained to recognize the condition of next NTS by extracting and analyzing shallow and abstract features from input historical data in this module. The network condition is then used to drive ABR decision module to choose the matching ABR model for rate adaptation. The framework of proposed neural network is shown in Fig.~\ref{figure:4}.

 The backbone is a 1D-CNN which is suitable to tackle with high-dimensional network throughput data. The input of neural network is historical NTS and the output is the recognition result indicating the network condition of next NTS. In the neural network, three convolutional layers with same architecture but different hyper-parameters including size of convolutional kernel and number of output channels, are devised to extract hierarchical features. In order to improve the capability of feature extraction and the accuracy of condition recognition, we add several optimized operations to the backbone network.1) We devise a multiple-perceptual-field mechanism by using multi-scale convolutional kernels in each convolutional layer to extract multi-scale features in the network throughput data. Specifically, we use three types of convolutional scales in each layer including ${\rm{3}} \times {\rm{1}}$, ${\rm{5}} \times {\rm{1}}$, and ${\rm{7}} \times {\rm{1}}$ kernels and then concatenate features after different scales of convolution operation in channel dimension. 2) We add Squeeze-and-Excitation (SE) module~\cite{Hu_2018_CVPR} in the network backbone to assign different weights for corresponding channels based on their degrees of contribution for final results. The SE operation is conducted through 1D-global pooling, fully connection and non-linear activation functions (Rectified Linear Unit and Sigmoid Function). After SE operation, weights of each channel are obtained and these values are multiplied with corresponding channels in original signal to achieve channel weighting. 3) We add residual structure~\cite{He_2016_CVPR} by transmitting shallow features to deeper layers directly and add shallow features with abstract features. It has been proved to address the problem of shallow-layer feature submerging and gradient vanishment/explosion when CNN becomes to be deeper. 4) We adopt channel shuffle operation to disturb original order of concatenated feature channels, which are obtained through various scales of convolution operation, for better generalization capability of the neural model. Feature channels are firstly divided into three groups equally. Then feature matrix is reshaped, transposed and reshaped to make feature channels shuffled. 5) We also perform the mean-std standardization for input throughput data and the batch normalization for data between layers at sample level. With these two normalization operations, the training process can be accelerated and neural network can converge to optimization in fewer epochs. What's more, dropout is used in fully-connection layers to avoid over-fitting in the training process.
 \vspace{-2mm}

\subsection{Condition-wised ABR Decision Module}\label{ssec:abr_decision}
\begin{figure}[htbp]
\centering
\includegraphics[width=3.5in]{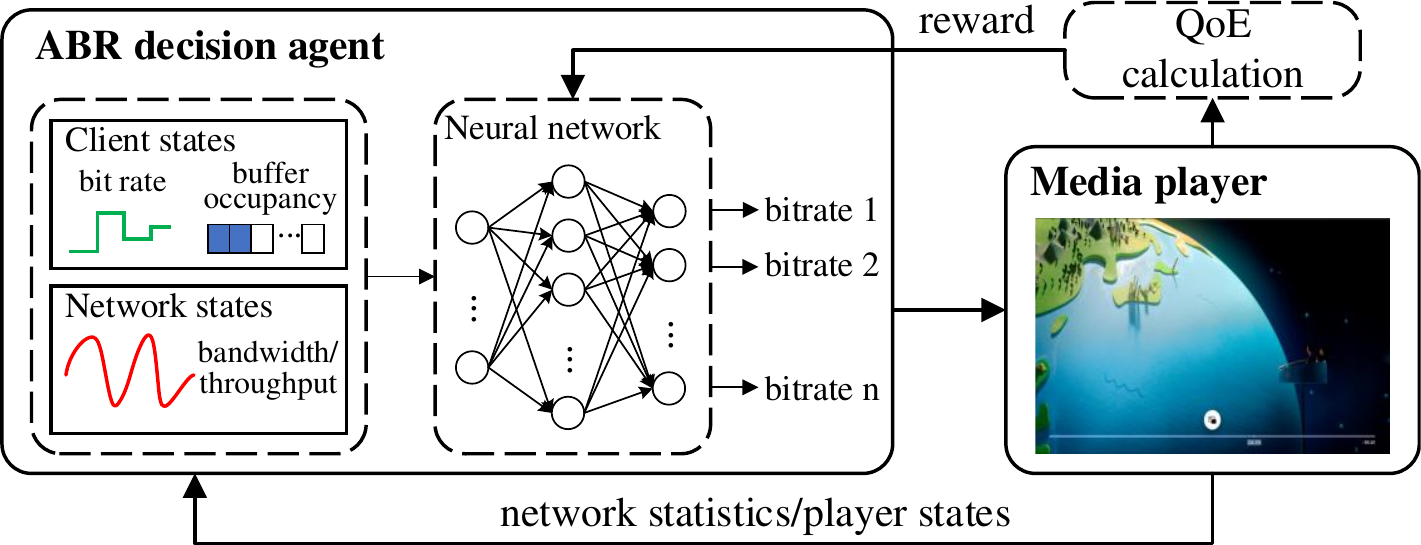}
\caption{Architecture of condition-wised ABR decision module.}
\label{figure:5}
\vspace{-2mm}
\end{figure}

 The condition-wised ABR decision module is constructed with a deep reinforcement-learning (RL) neural network. It is deployed to make bit rate decisions using one of trained ABR models indicated by condition learning module mentioned above. The goal of ABR decision module is to achieve the maximum reward signal, which reflects video QoE of users. In the training phase, the agent obtains a variety of observations from the video streaming environment, including network statistics (i,e., bandwidth or throughput) and player states at client-side (i.e., buffer occupancy, bit rate and so on), and inputs them as states into the RL neural network. Neural network produces actions based on these input states. Afterward, decisions made by neural network will change the states and get corresponding reward from the environment. The RL neural network is trained with different states to get a higher video QoE of users in the future. When the training reaches convergence, the neural network can enter into working phase which makes decisions based on observations of environment. The architecture of condition-wised ABR decision module is shown as Fig.~\ref{figure:5}.

We conduct the training phase under various network conditions in order to generate several ABR models. Every once in a while, certain ABR model corresponding to certain network condition is selected to make bit rate decision according to recognition results of condition learning module.


The algorithm of RL in ABR decision module we use is Asynchronous Advantage Actor-Critic (A3C)~\cite{pmlr-v48-mniha16}, a state-of-the-art actor-critic method. The RL neural network consists of two parts: an actor network, which is used to make bit rate decisions for each video chunk, and a critic network, which is used to evaluate the performance of decisions made by the actor network. The inputs adopted for reinforcement learning agent (RL-agent)  include historical throughput, download time, future chunk size, current buffer size, last chunk bit rate and number of chunks left. The policy gradient method is adopted to train the critic network and the actor network. The RL-agent takes an action ${a_t}$ that corresponds to the bit rate for the next chunk when receiving a ${s_t}$. After applying each action to client-side player, the environment will provide the RL-agent with a reward ${r_t}$ for that video chunk. In this paper, the reward is set according to the QoE metric proposed in MPC~\cite{yin2015control} to reflect the performance of each chunk download. The primary goal of the RL-agent is to maximize the expected cumulative reward received from the environment.

\section{Implementation of {\it ANT}}
\subsection{Introduction of Network Traces}
Since it's time-consuming to ``experience'' video downloads in the real-world streaming environment, we use simulations over a wide range of network traces in the training and testing phases, including traces collected from public dataset (broadband  dataset provided by FCC, 3G/HSDPA mobile dataset collected in Norway, 4G/LTE bandwidth from Belgium, traces from Oboe and traces from ACM Live Streaming platform) and traces collected online from Tencent video platform (Wifi network traces and 3G/4G network traces). In the network trace files, time information (second) and corresponding throughput (bandwidth) information (Mbit per second) are contained. We randomly divide all {\it 2658} traces into training set and testing set respectively by the proportion of 80\% and 20\%. NTS in 20-second duration are provided for trace aggregation and the classified threshold mentioned above is set as $2/3$. Correspondingly, throughput learning will be conducted every 20 seconds.

\subsection{Design of Confidence Mechanism}
\begin{figure}[htbp]
\centering
\includegraphics[width=3.5in]{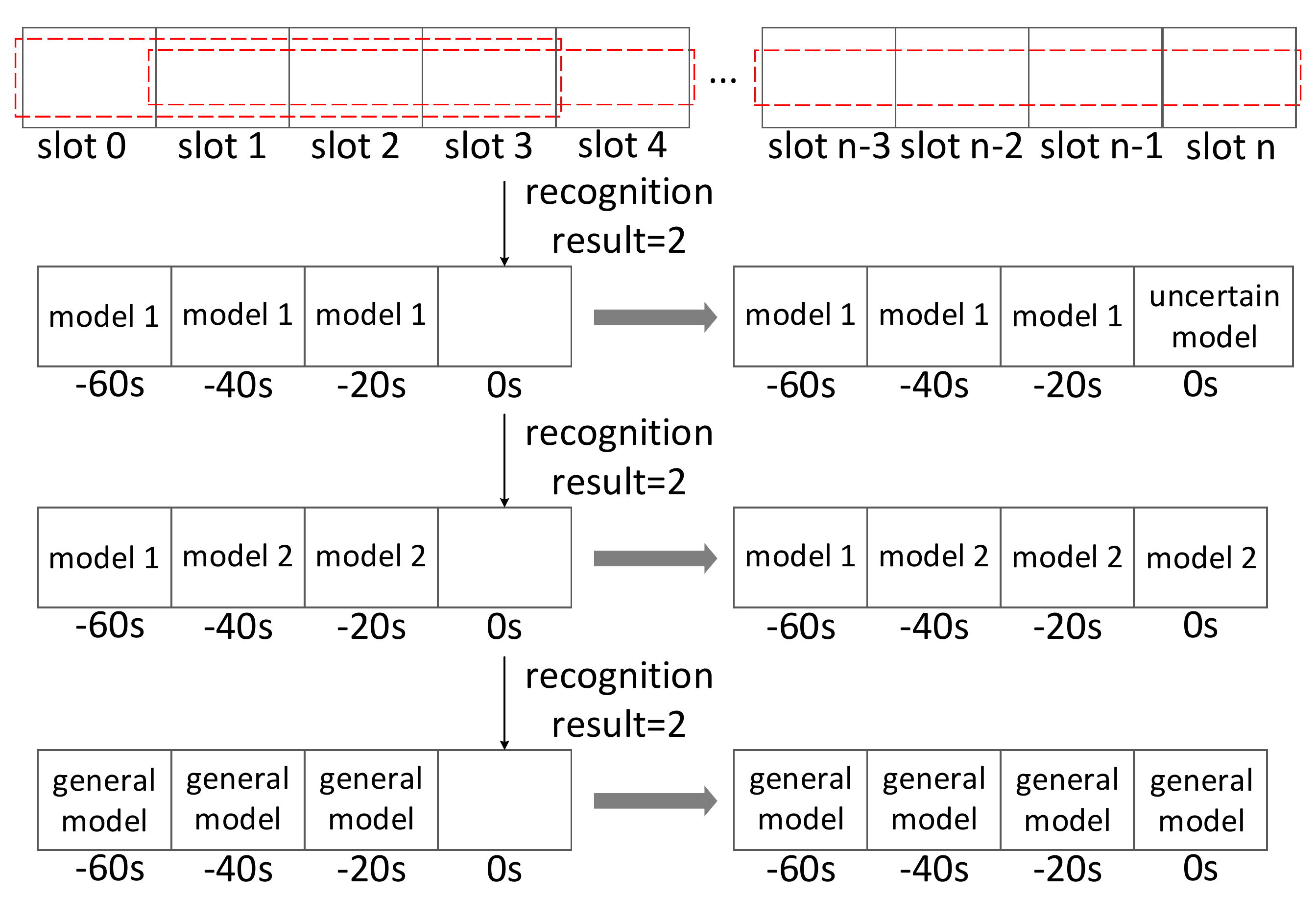}
\caption{Illustration of confidence mechanism.}
\label{figure:6}
\vspace{-2mm}
\end{figure}

In the trace aggregation mechanism, there may be some traces of uncertain condition due to not reaching the threshold, as shown in Fig.~\ref{figure:3}. In order to tackle with these inaccurate recognitions, we train a general ABR model under all various network traces, and a dedicated ``uncertain'' model under the traces of uncertain condition type. To be specific, when condition learning model outputs a different result compared with that in last period, we would not conduct model switching immediately. This is because that the difference may be incurred by inaccurate recognition results (as the recognition accuracy is not 100\%, see Sec.~\ref{ssec:condition_recognition}) instead of real network condition switching. In this case, switching the ABR model may incur inappropriate model selection, which may significantly degrade the ABR performance. Thus, we propose a confidence mechanism based on the sliding window in implementation phase. Condition learning is conducted  every 20 seconds and recognition result at each step is queued into a sliding window. Current result is admitted only if the decision is equal to past two out of three decisions. Otherwise, the ``uncertain'' ABR model will be selected for the following 20-second NTS's ABR decision. In addition, when the video streaming system runs in the initial period (i.e., 60 seconds in the beginning) that not enough historical throughput data to conduct condition learning, the general ABR model is selected until the condition of confidence mechanism is met. The confidence mechanism is illustrated in Fig.~\ref{figure:6}.

\subsection{More Details of Implementation}
 We implement the RL-based ABR decision module and CNN-based condition learning module using Tensorflow. Bit rates of video content we choose include 135Kbps, 340Kbps, 835Kbps, 1350Kbps and 2640Kbps. The actor network learning rate is set as 0.0001 and critic network learning rate is set as 0.001. The entropy weight of actor network is set as 0.5. Actor network and critic network both include 1 hidden layer with 1D-CNN network with the kernel size as 4 and 128 output channels, and a fully-connection network with 128 neurons. The number of RL-agent is set as 16. We consider past 8 states observed from the environment, which are normalized before input into ABR agents. Rebuffer penalty and smoothness penalty are set as 2.64 and 1 respectively. For neural network in condition learning module, three types of CNN filters are ${\rm{3}} \times {\rm{1}}$, ${\rm{5}} \times {\rm{1}}$, and ${\rm{7}} \times {\rm{1}}$ respectively. The number of output channels of each CNN layer is ${\rm{64}} \times {\rm{3}}$, ${\rm{128}} \times {\rm{3}}$, and ${\rm{256}} \times {\rm{3}}$ respectively. Kernel size in pooling layers we choose is ${\rm{2}} \times {\rm{1}}$. The first layer of fully-connection network has 256 neurons and 128 neurons are contained in the second layer. For all convolution operations and pooling operations, we set stride with 1 and add padding operation to maintain the data width. As for other hyper-parameters, We set learning rate as 0.0001 and batch size as 80 in the training phase.
 The device we use to train and test neural networks is a Ubuntu 16.04 server with Intel Xeon CPU E5-2683 v4 @ 2.10GHz and Nvidia GeForce GTX 1080Ti 11G GPU.

\section{Experiment Results and Analysis}
\subsection{Network Trace Clustering}
\begin{figure}[htbp]
    \centering
    \subfigure[Trend of SSE with $k$ Values]{  \includegraphics[width=0.48\linewidth]{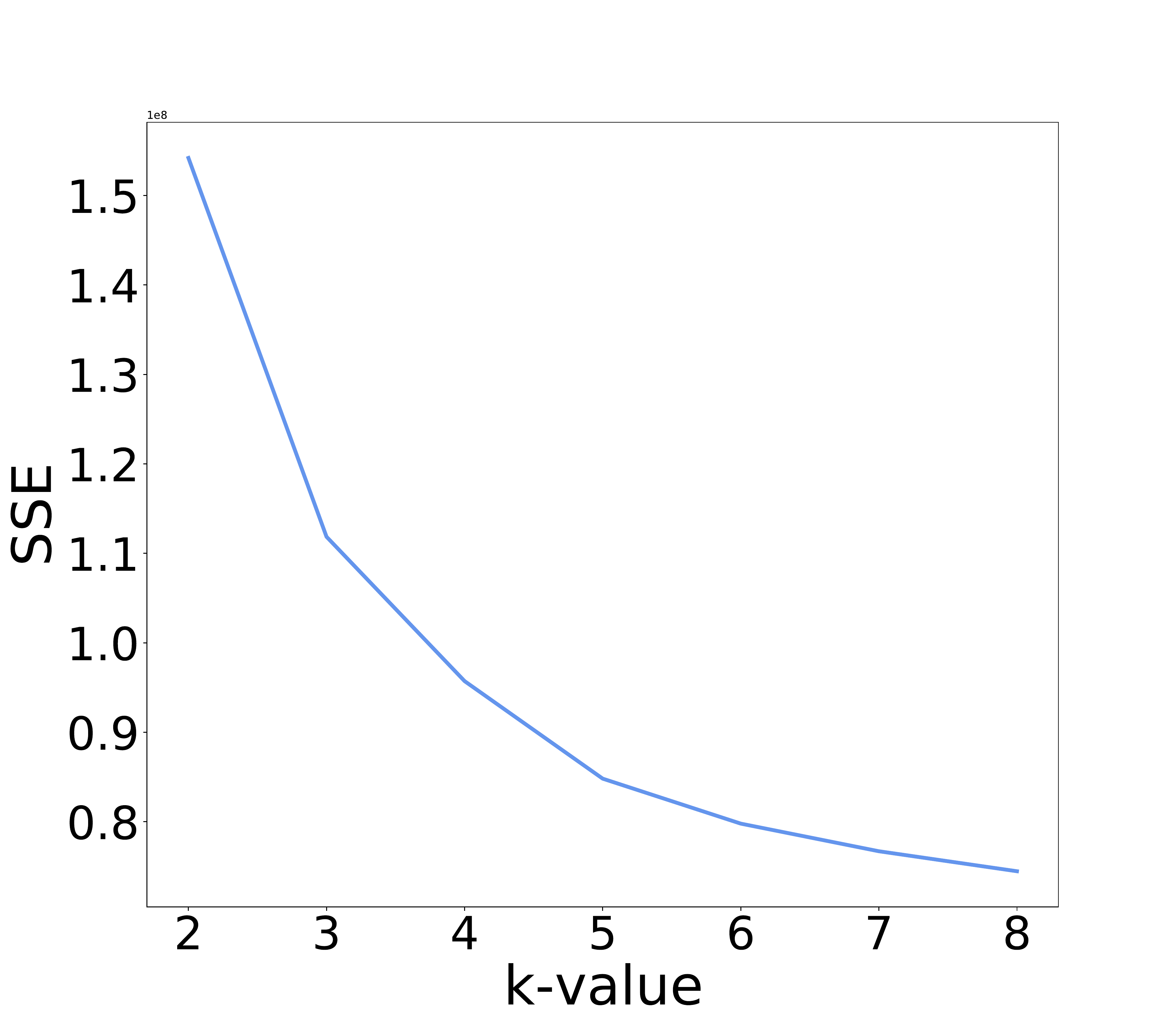} \label{a}}\hspace{-1mm}
    \subfigure[Trend of DBI with $k$ Values]{  \includegraphics[width=0.48\linewidth]{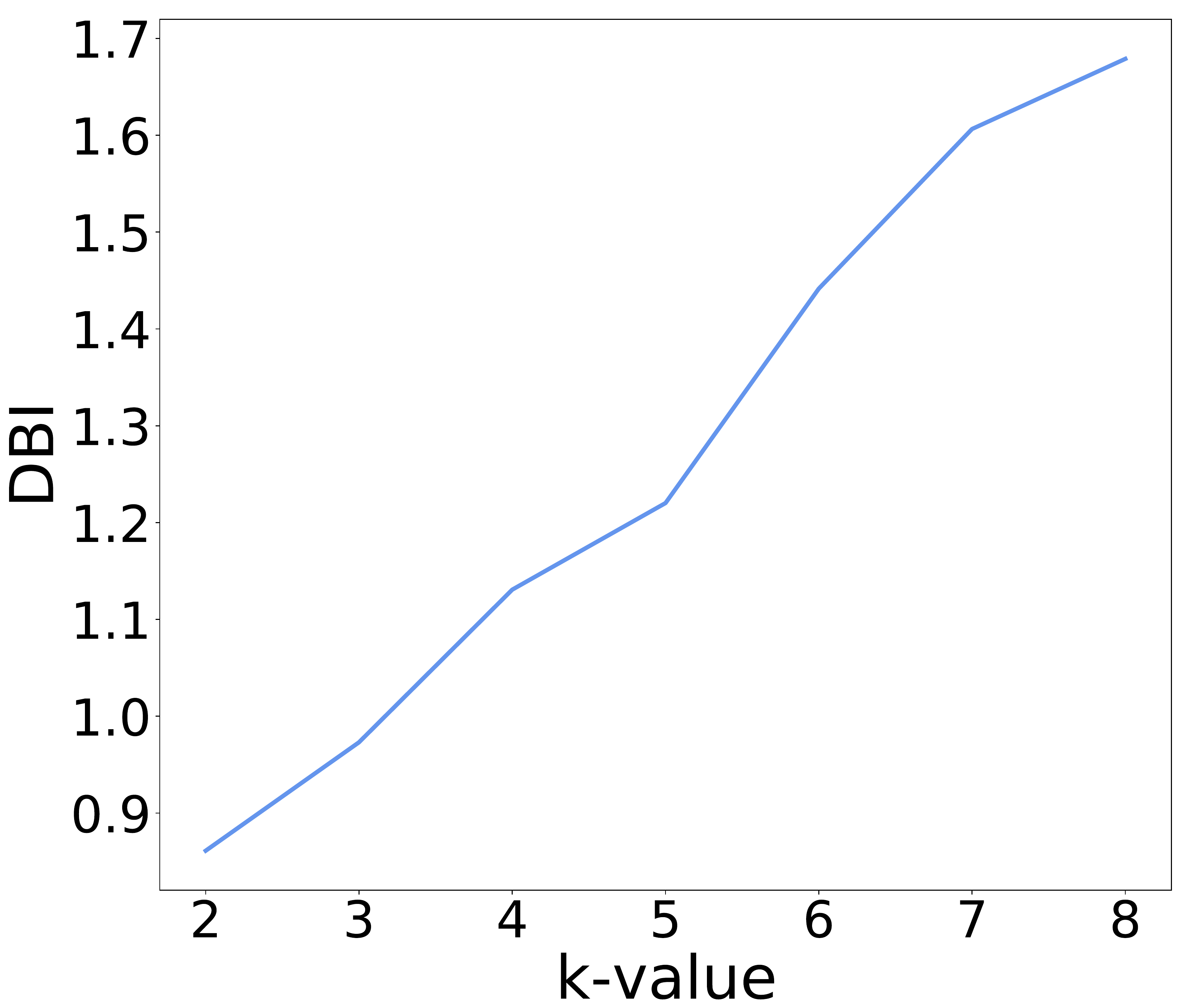} \label{b}}
    \caption{Trends of indicators with $k$ values.}
    \label{figure:7}
\vspace{-1mm}
\end{figure}

Before training various ABR models under network traces of different conditions, traces are split into 20-second segments and feature aggregation is conducted in the segment dimension as mentioned in Sec.\ref{ssec:condition_learning}. The number of clusters $k$ is critical when running K-means algorithm. To determine the most appropriate value $k$ considering all segments, we conduct K-means clustering with parameter $k$ varying from 2 to 8, and then make comparison under the metrics of Sum of the Squared Errors (SSE) and Davies-Bouldin Index (DBI). The trends of indicators with $k$ values are shown in Fig.~\ref{figure:7}.

The SSE value becomes smaller gradually with the number of clusters $k$ increasing. When $k$ approaches the most appropriate value, the downward trend (slop) will become slower till convergence. On the contrary, DBI calculates the ratio of the degree of separation between clusters to the degree of aggregation within a cluster. Therefore, DBI value increases gradually with the number of clusters increasing till convergence. In the view of these two indicators, we find the turning point where trend slows down occurs at $k=5$. Thus, we set the number of clusters as 5 for trace segments, and 6 for whole traces considering extra ``uncertain'' condition after applying proposed confidence mechanism. 

\subsection{Network Condition Recognition} \label{ssec:condition_recognition}
In order to recognize network conditions based on historical throughput data, we use clustered trace segments to train condition learning module. The length of input historical throughput data we adopt is 20 seconds. We choose 80\% and 20\% of all throughput data respectively as training data and validation data. We conduct the training process until the validation accuracy reaches convergence. After 100 epochs, the validation accuracy converges to 98.56\%. It proves that throughput data in 20-second duration contains adequate information, based on which condition learning module could recognize almost all the network conditions accurately.

With the capability of recognizing network condition accurately, and the assistance of confidence mechanism, the network condition learning module can drive ABR model switching for better bit rate decision based on historical throughput data.

\subsection{ABR Decision and Final QoE}
\begin{figure}[htbp]
    \centering
    \subfigure{  \includegraphics[width=0.465\linewidth]{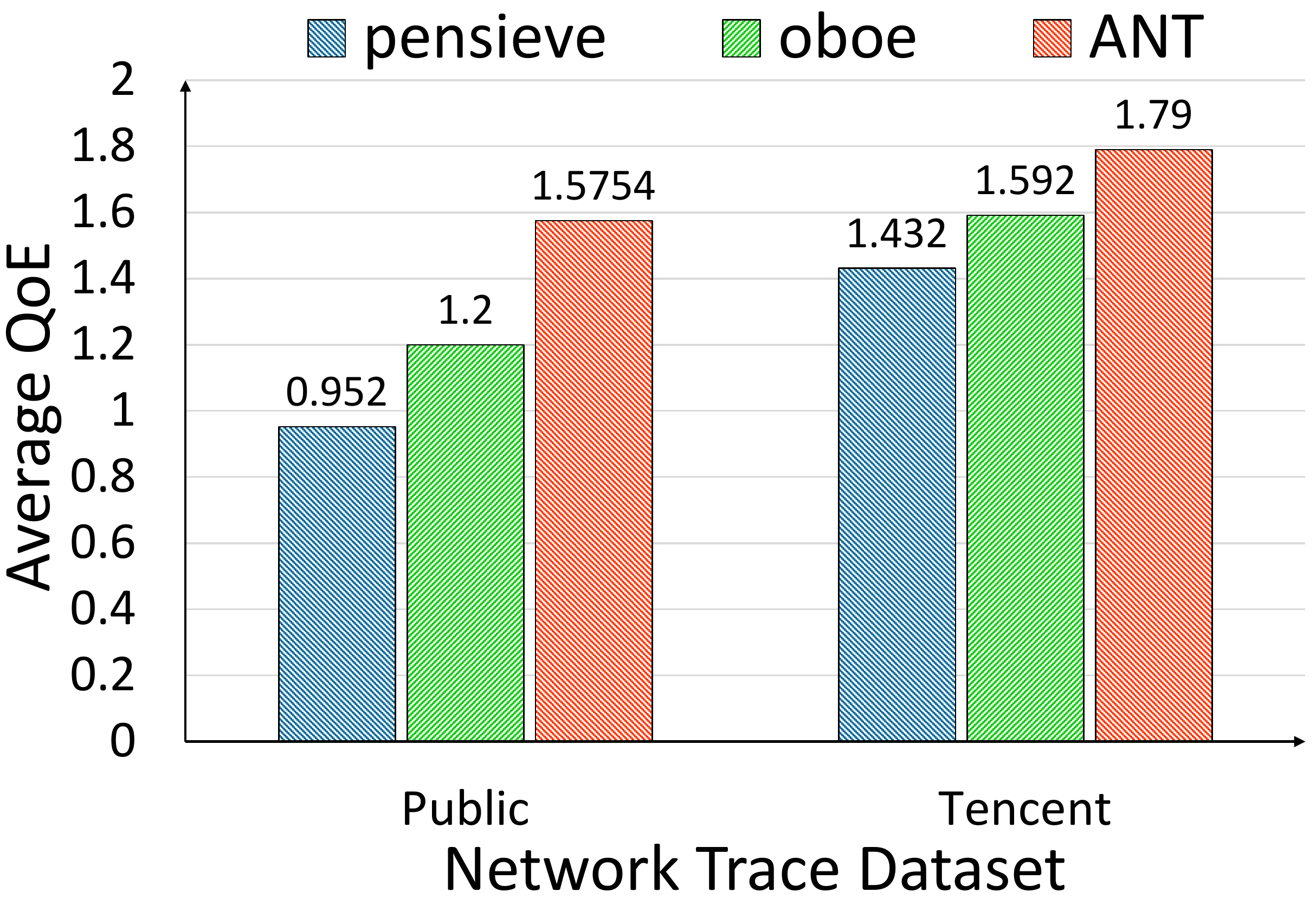} }
    \subfigure{  \includegraphics[width=0.465\linewidth]{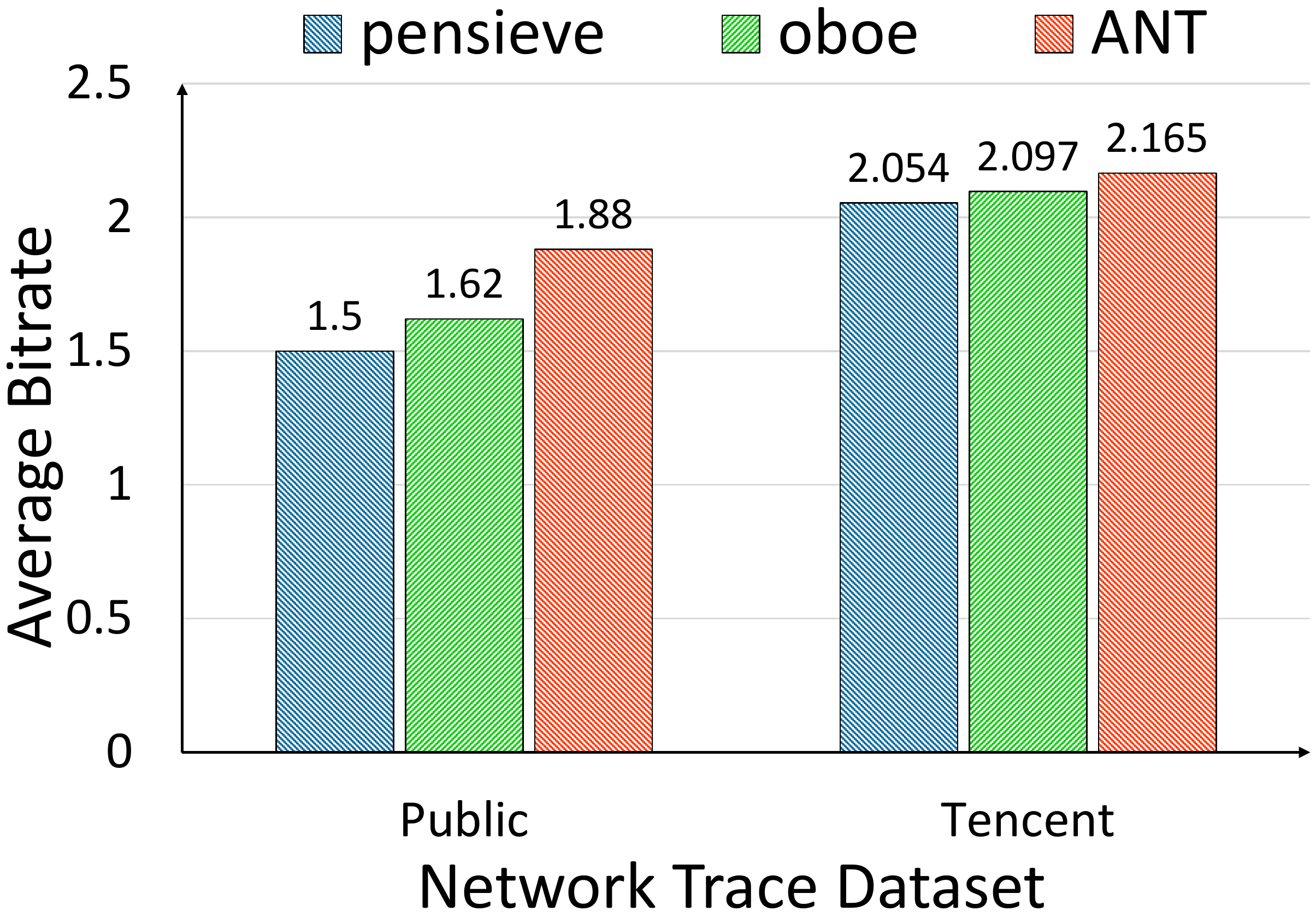} }
    \subfigure{  \includegraphics[width=0.464\linewidth]{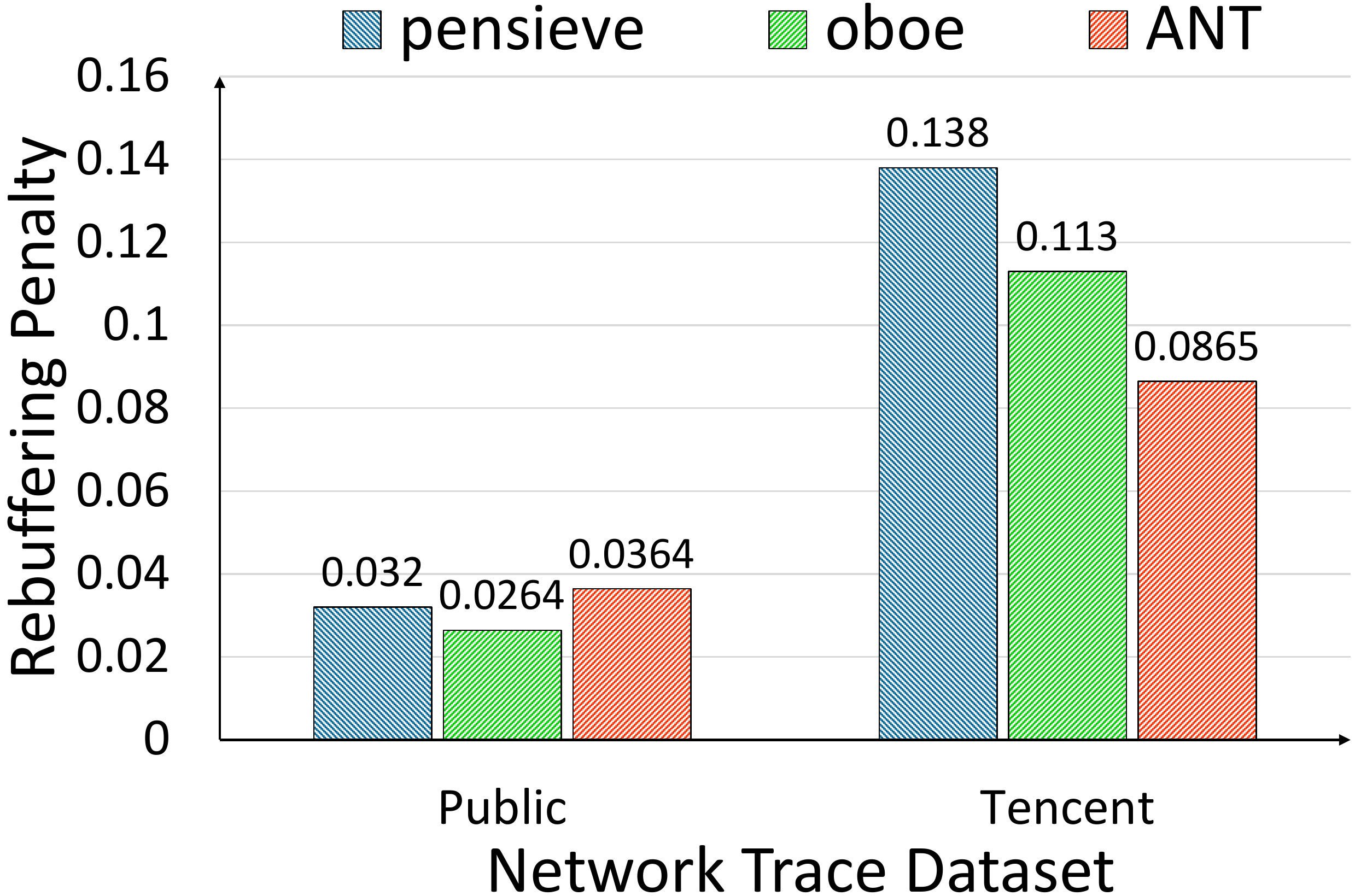} }
    \subfigure{  \includegraphics[width=0.464\linewidth]{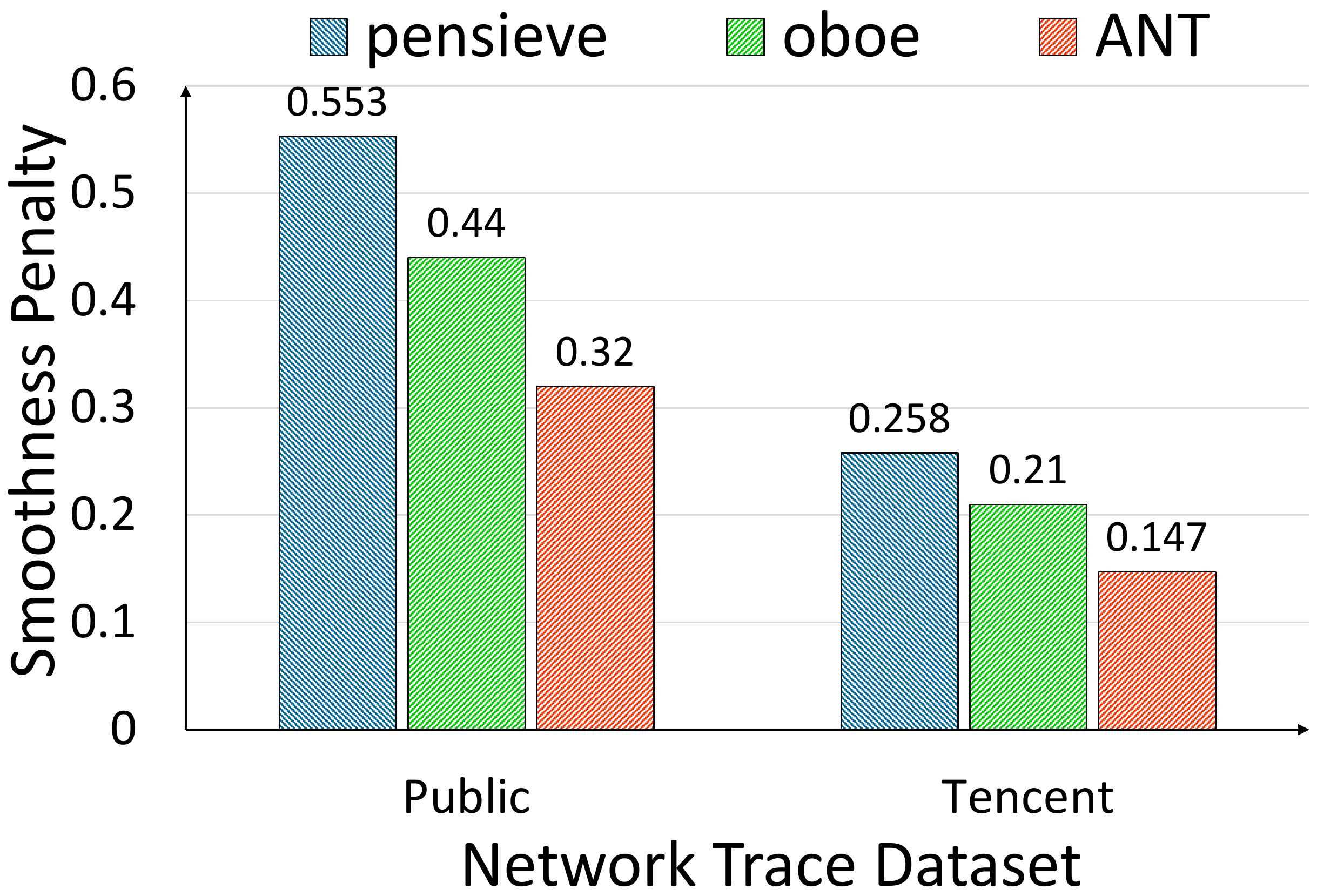} }
    \caption{Final performance comparison under both public traces and Tencent traces.}
    \label{figure:8}
\end{figure}

\begin{figure}[htbp]
    \centering
    \subfigure{  \includegraphics[width=0.47\linewidth]{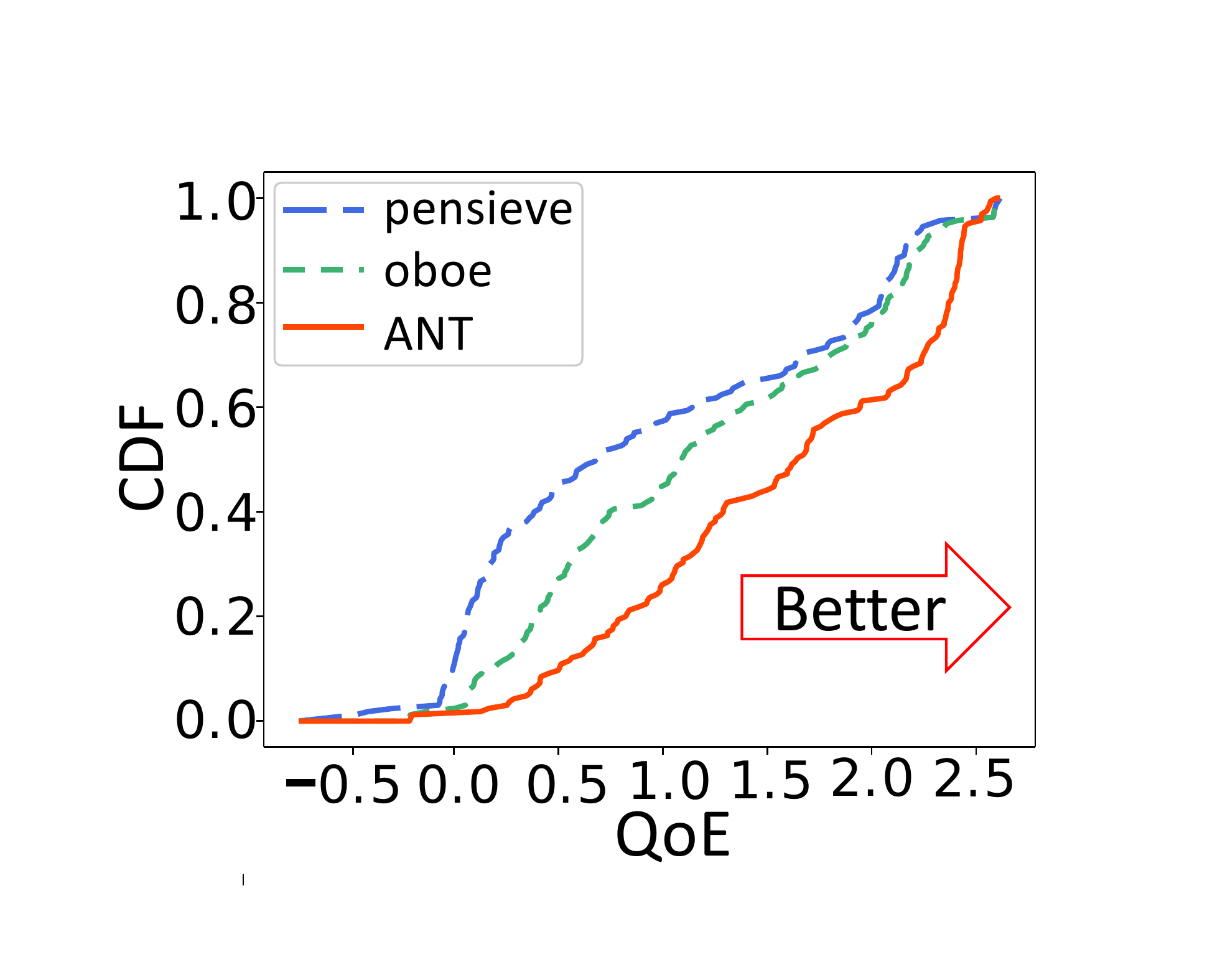} }\hspace{-1mm}
    \subfigure{  \includegraphics[width=0.47\linewidth]{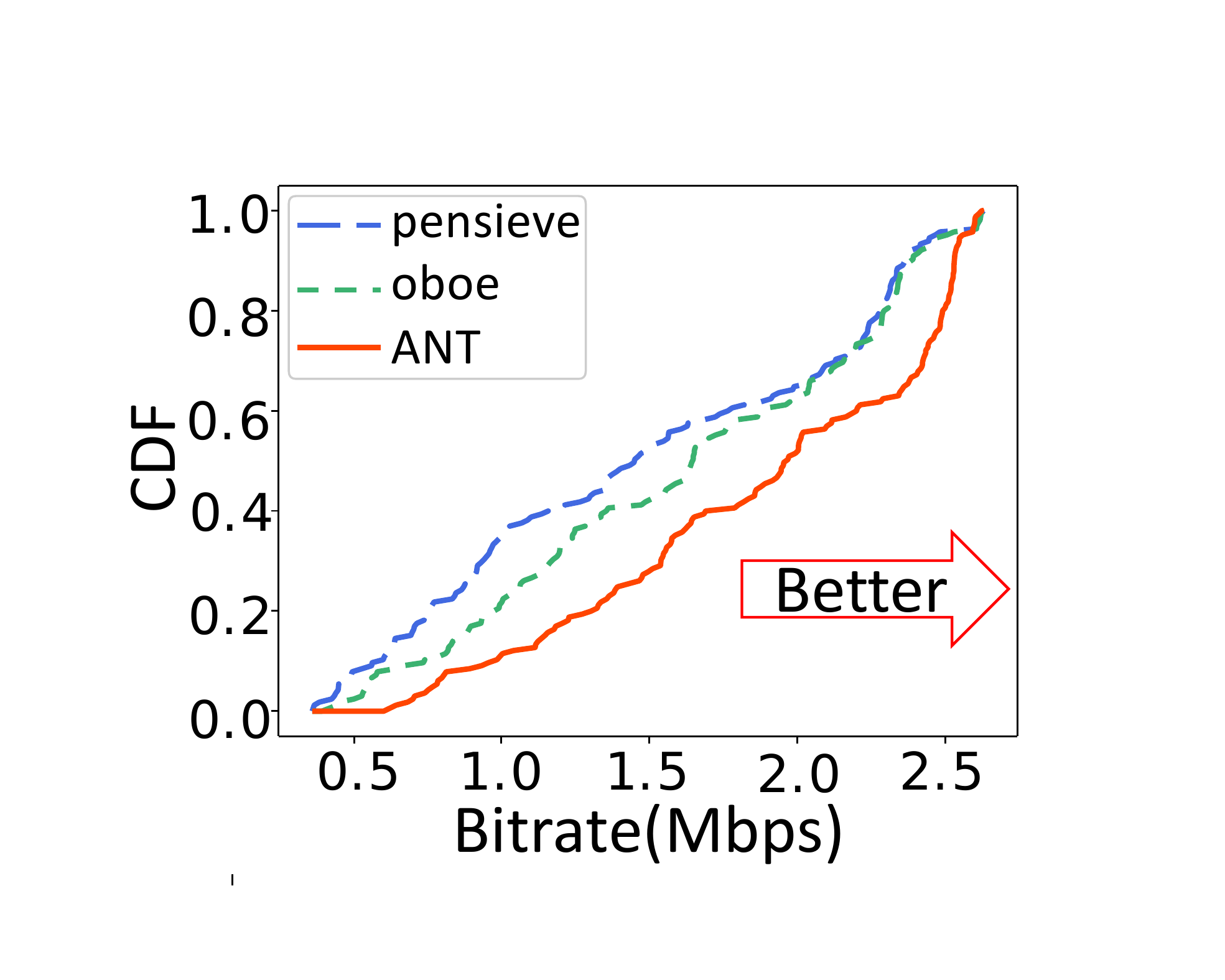} }
    \subfigure{  \includegraphics[width=0.47\linewidth]{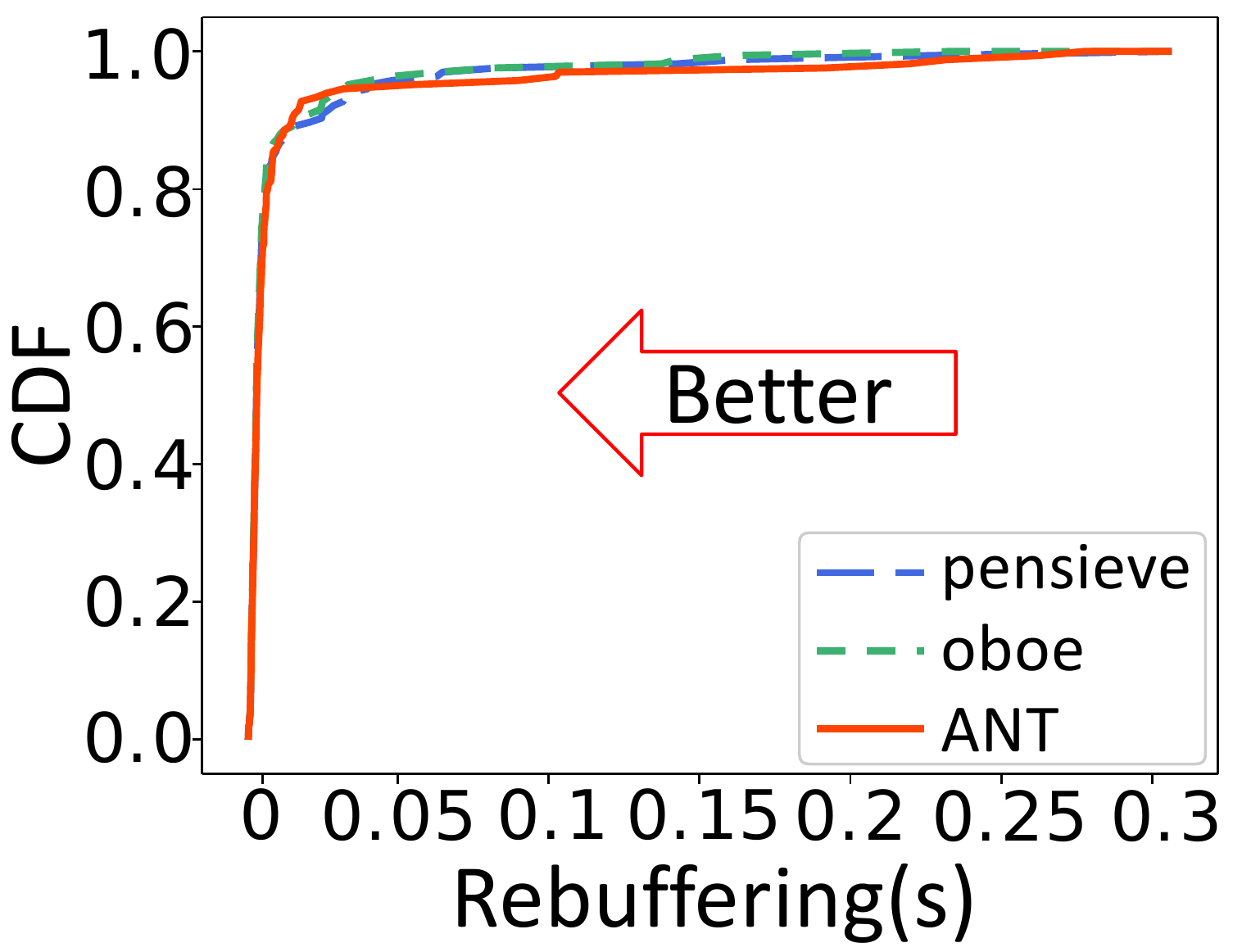} }\hspace{-1mm}
    \subfigure{  \includegraphics[width=0.47\linewidth]{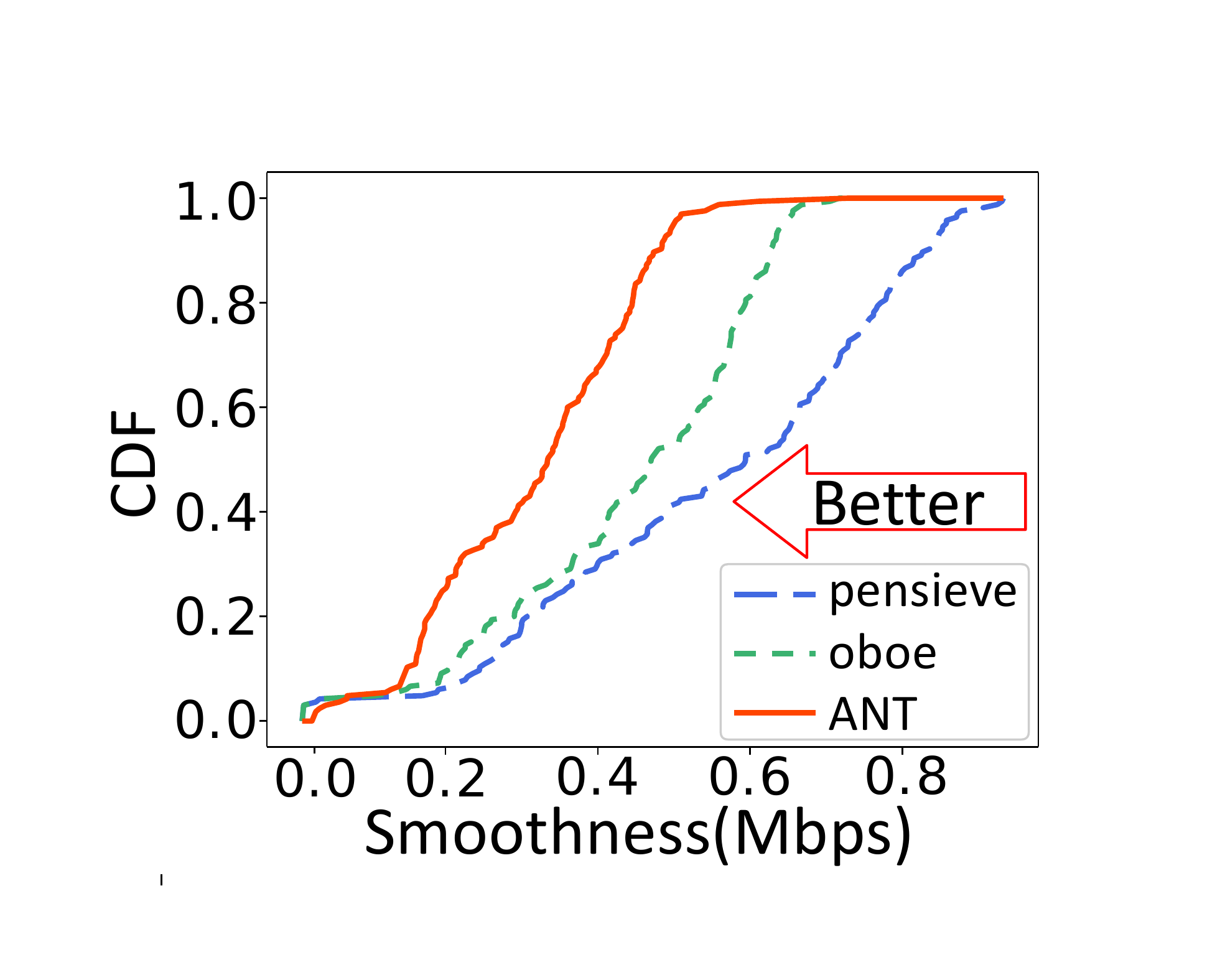} }
    \caption{Final CDF curve under public traces.}
    \label{figure:9}
    \vspace{-1mm}
\end{figure}

\begin{figure}[htbp]
    \centering
    \subfigure{  \includegraphics[width=0.47\linewidth]{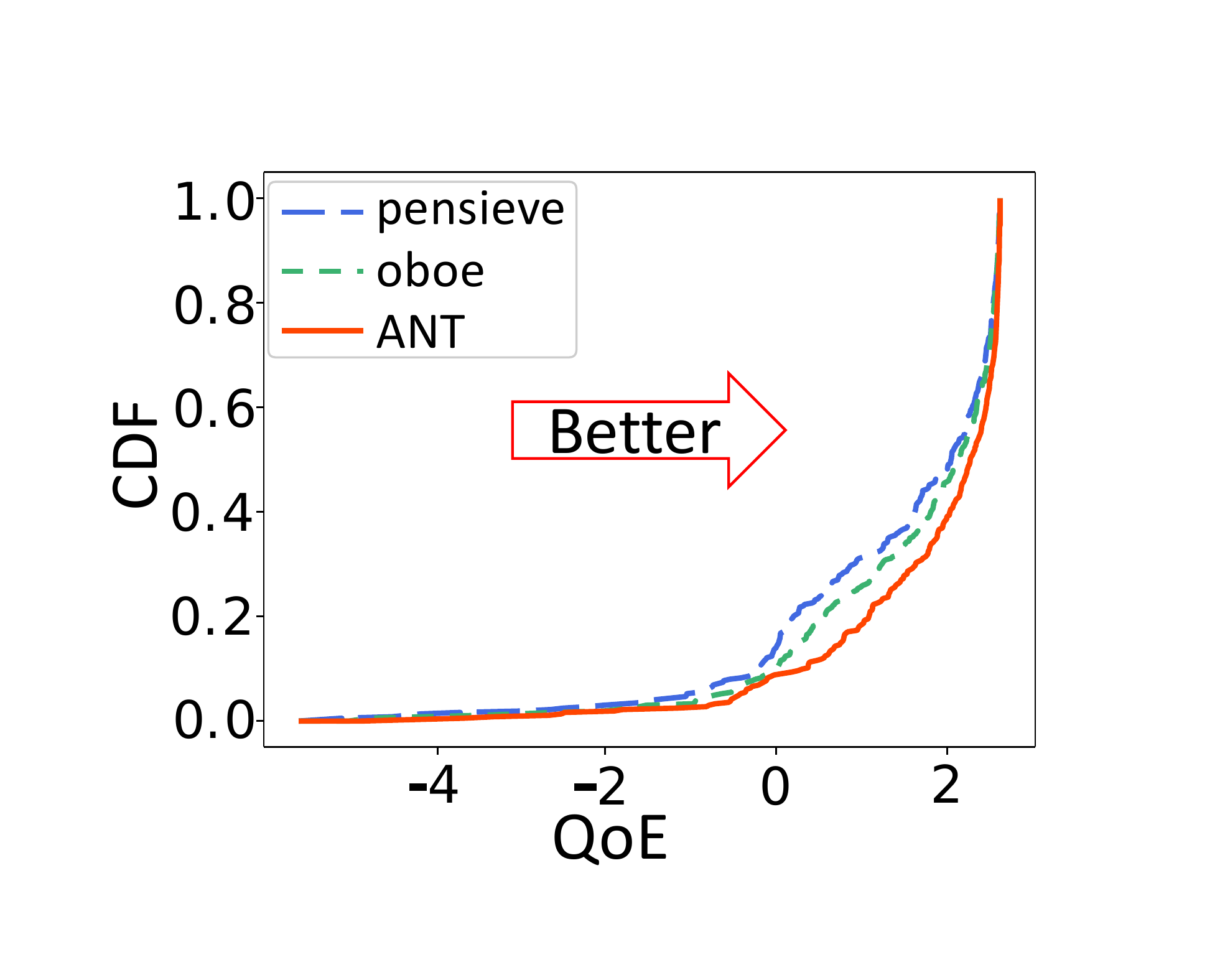} }\hspace{-1mm}
    \subfigure{  \includegraphics[width=0.47\linewidth]{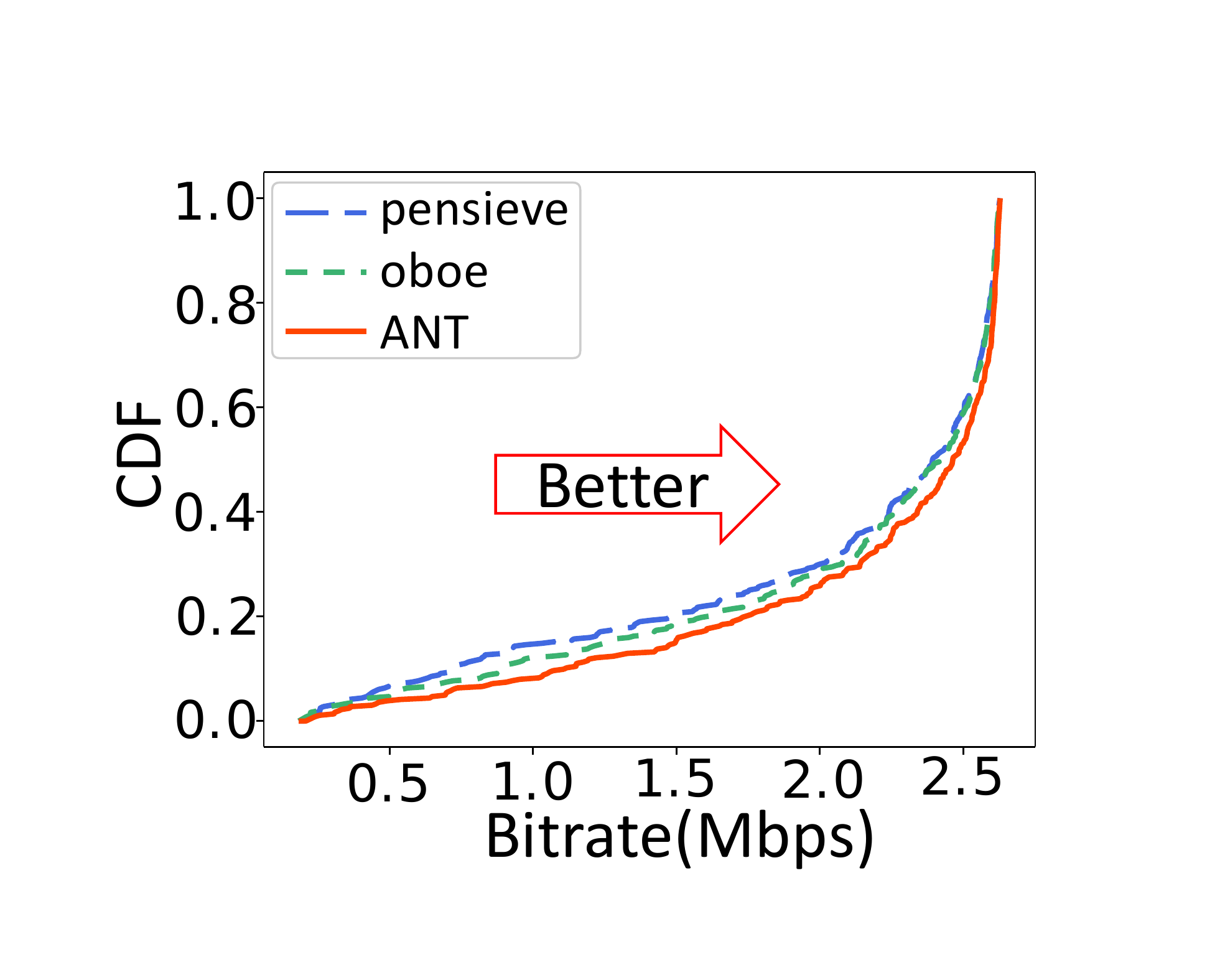} }
    \subfigure{  \includegraphics[width=0.47\linewidth]{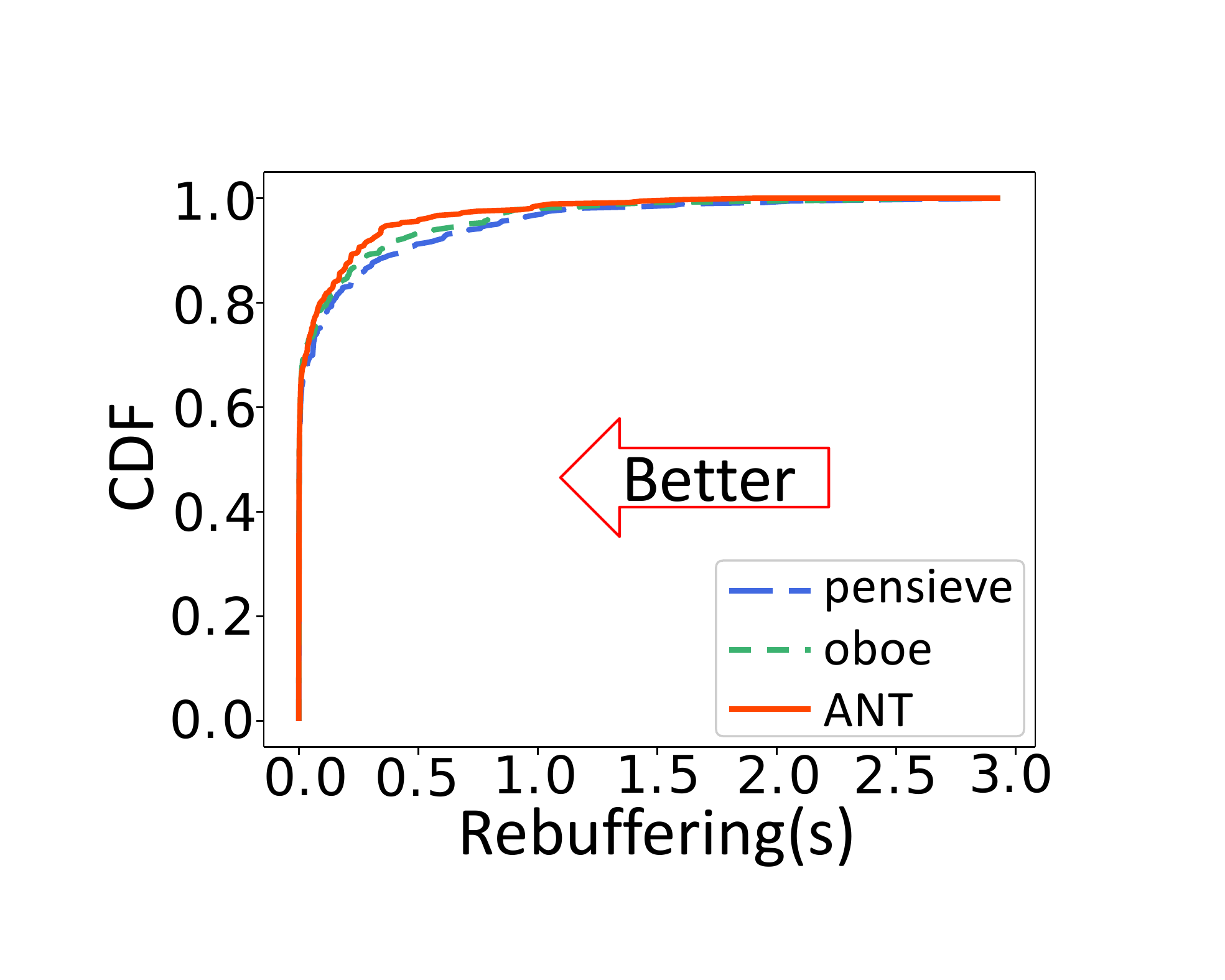} }\hspace{-1mm}
    \subfigure{  \includegraphics[width=0.47\linewidth]{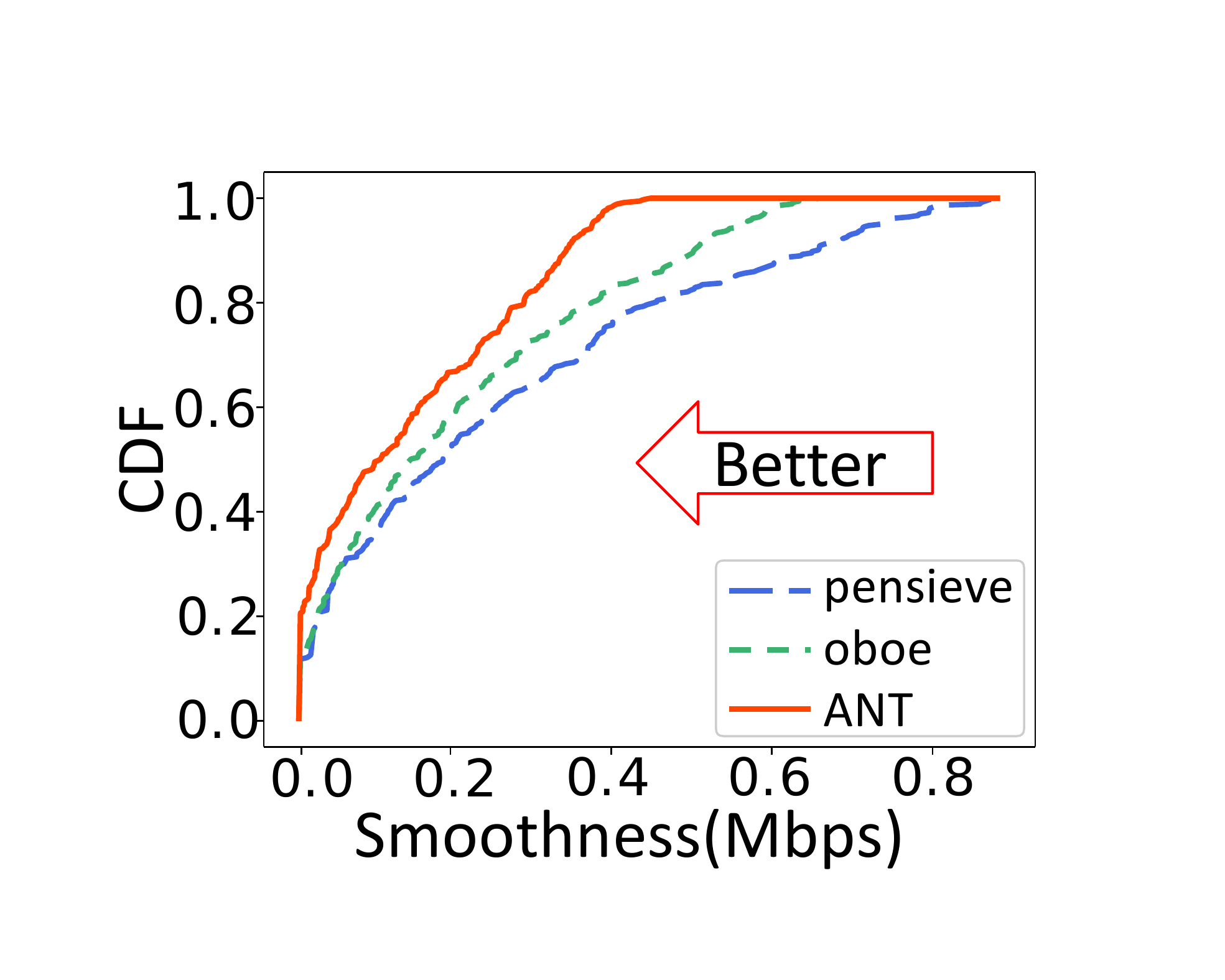} }
    \caption{Final CDF curve under Tencent traces.}
    \label{figure:10}
    \vspace{-1mm}
\end{figure}

In this part, we evaluate the performance of {\it ANT} for bit rate adaptation on the QoE metric and its individual components, including bit rate utility (Mbps), rebuffering penalty (seconds) and smoothness penalty (Mbps), under diverse network traces. Two state-of-the-art ABR algorithms, i.e., Pensieve, a RL-based ABR algorithm with one general model, and Oboe, an ABR algorithm with auto-tuning mechanism according to network conditions, are adopted for performance comparing. For the Oboe algorithm, we divide all network traces into 5 parts according to the average value of throughput, i.e., 0-3Mbps, 3-6Mbps, 6-9Mbps, 9-12Mbps and over 12Mbps. Then a similar training (to Sec.~\ref{ssec:abr_decision}) is conducted to generate ABR models corresponding to throughput conditions. In addition, we evaluate their performance under another set of network traces collected from Tencent video platform. The results are shown in Fig.~\ref{figure:8} to Fig.~\ref{figure:10}.

As shown in Fig.~\ref{figure:8}, we calculate the average values of final QoE and its individual components for each chunk under all public traces. {\it ANT} achieves 1.575 on final QoE metric and 1.88Mbps in average bit rate for each video chunk which are 65.5\% and 25.3\% higher respectively than Pensieve algorithm. In addition, {\it ANT} rivals Pensieve in terms of average rebuffering time and achieves much lower average bit rate fluctuations (higher smoothness) than pensieve. On the other hand, {\it ANT} achieves 31.28\% and 16.05\% improvement in view of average QoE and average bit rate, compared with Oboe's auto-tuning mechanism. Similarly, {\it ANT} shows less bit rate fluctuations. Although Oboe performs slightly better than proposed {\it ANT} for average rebuffering time under traces from public platform, it is acceptable because of much better performance of {\it ANT} in view of final QoE, bit rate utility, and bit rate fluctuation.

For the evaluation under Tencent network traces, the results are shown in Fig.~\ref{figure:8}. {\it ANT} achieves 1.79 on final QoE metric and 2.165Mbps in average bit rate for each video chunk, which are 25\% higher and 5.4\% higher than Pensieve respectively. And in terms of average rebuffering time and average bit rate fluctuations, {\it ANT} outperforms Pensieve as well. Compared to Oboe's auto-tuning mechanism, {\it ANT} achieves 12.44\% and 3.24\% improvement in the form of average QoE and average bit rate, and less rebuffer time and bit rate fluctuations throughout the whole video streaming process.

We further calculate the CDF (Cumulative Distribution Function) of final QoE values and  corresponding individual components under both the public traces and Tencent traces, and the results are shown in Fig.~\ref{figure:9} and  Fig.~\ref{figure:10} respectively. CDF curves also illustrate consistent results. We can find that {\it ANT} achieves better and robust performance than existing state-of-the-art algorithms (i.e., Pensieve and Oboe) on various network conditions.

\subsection{Performance Stability Analysis}
\begin{figure}[htbp]
    \centering
    \subfigure[Public Dataset]{  \includegraphics[width=0.47\linewidth]{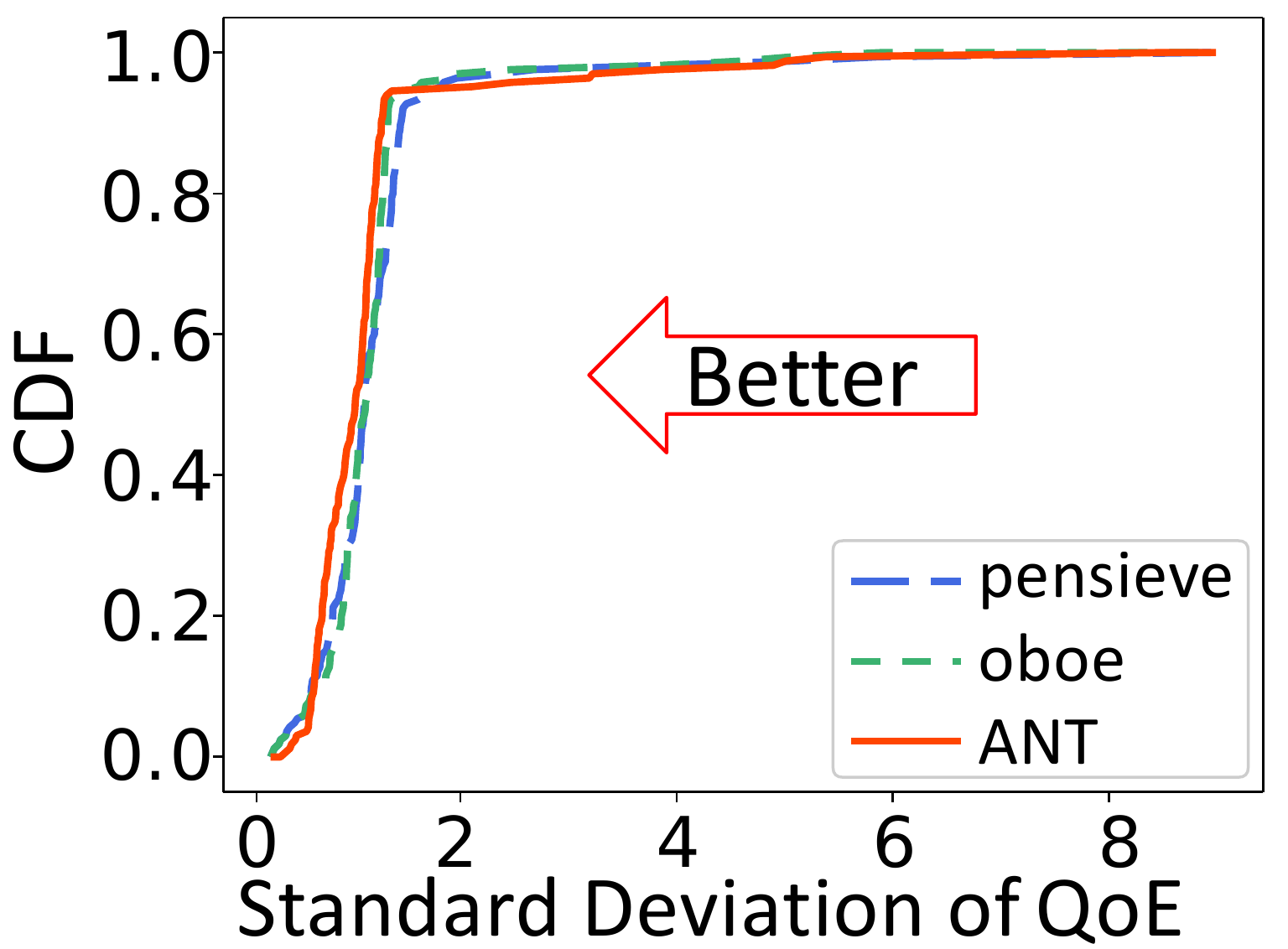}\label{a}}\hspace{-1mm}
    \subfigure[Tencent Dataset]{  \includegraphics[width=0.47\linewidth]{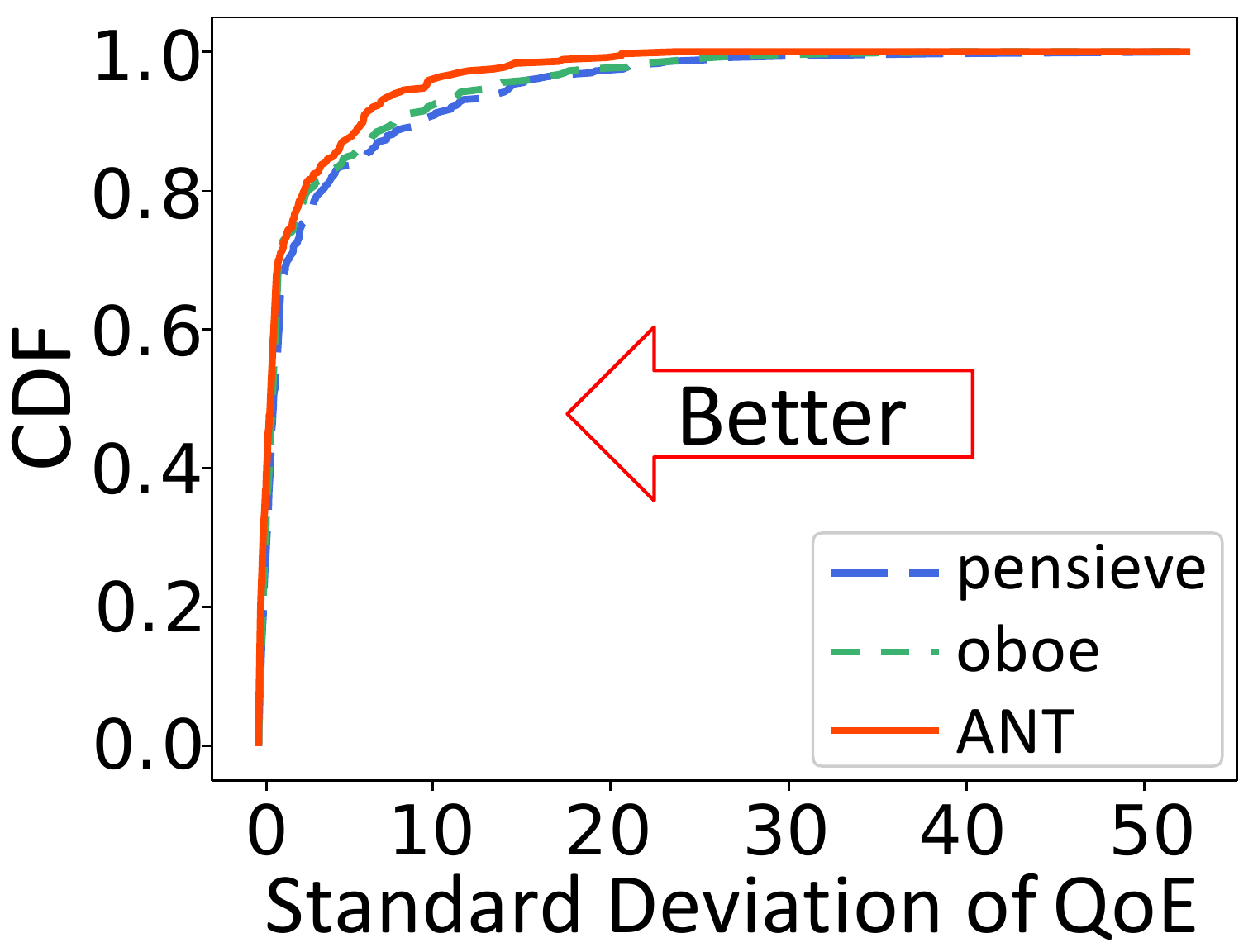}\label{b} }
    \caption{CDF curve for the standard deviation of QoE.}
    \label{figure:11}
\end{figure}

\begin{figure}[htbp!]
    \centering
    \subfigure{  \includegraphics[width=0.47\linewidth]{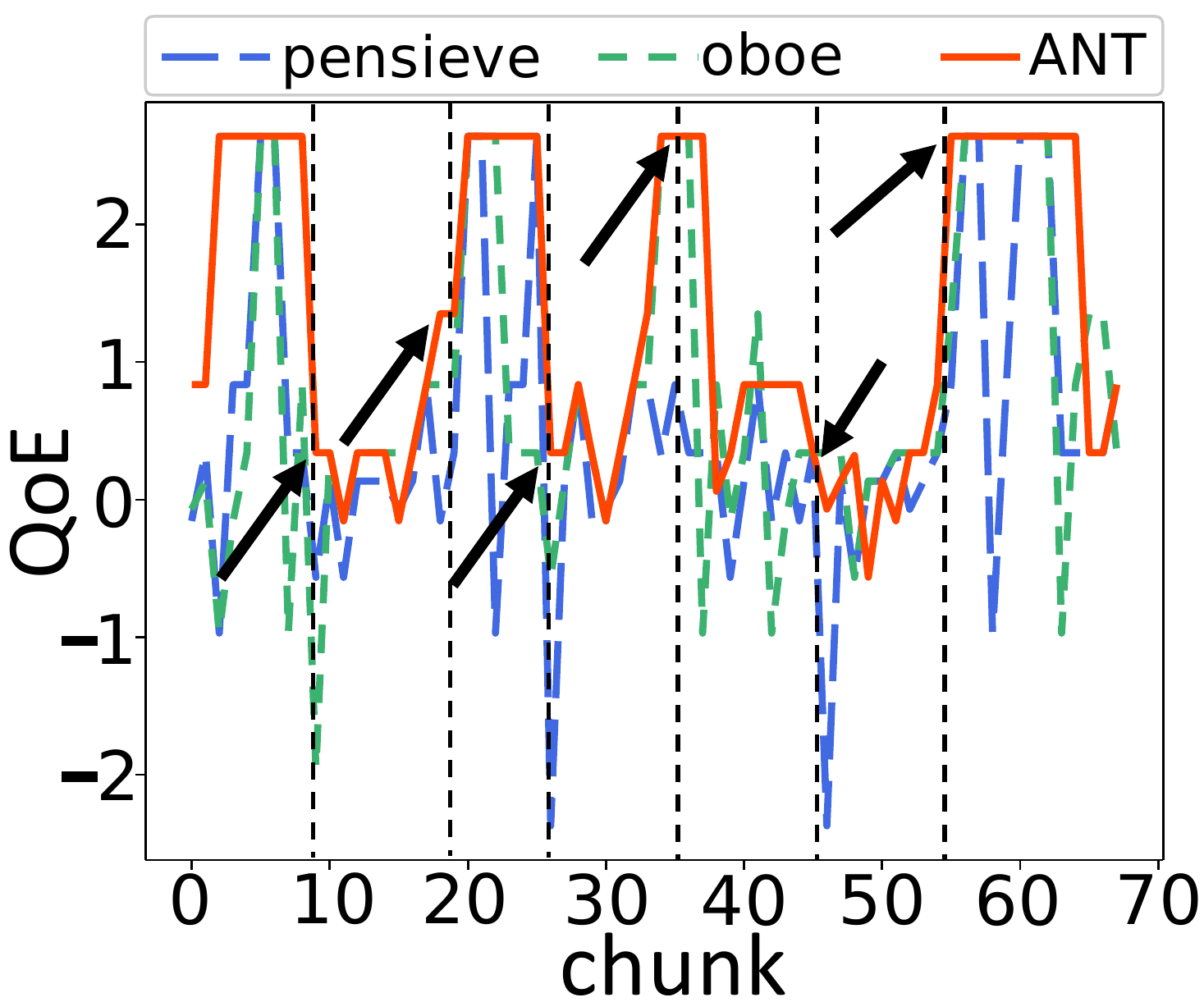} }\hspace{-1mm}\hspace{-1mm}
    \subfigure{  \includegraphics[width=0.47\linewidth]{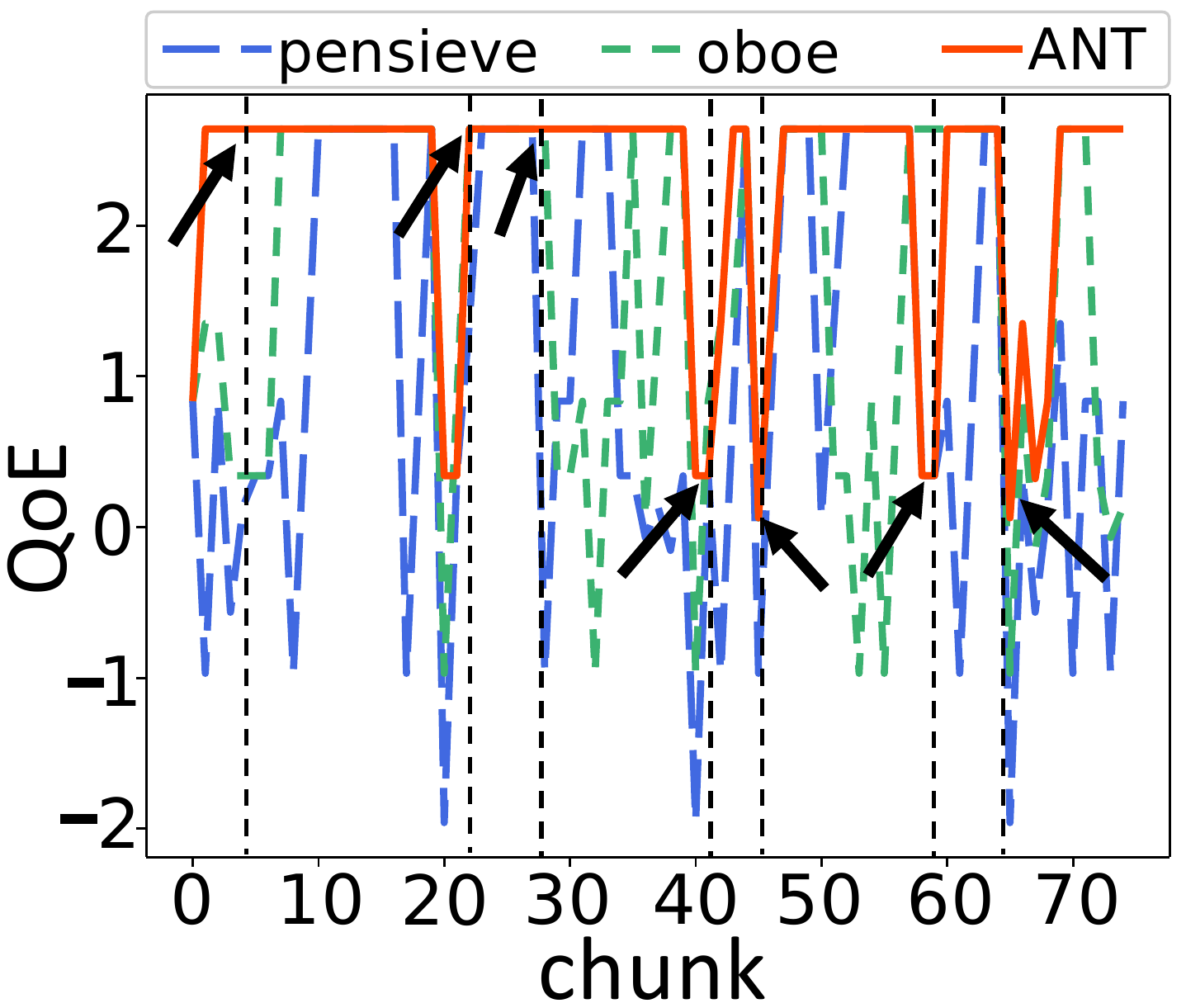}  }
    \subfigure{  \includegraphics[width=0.47\linewidth]{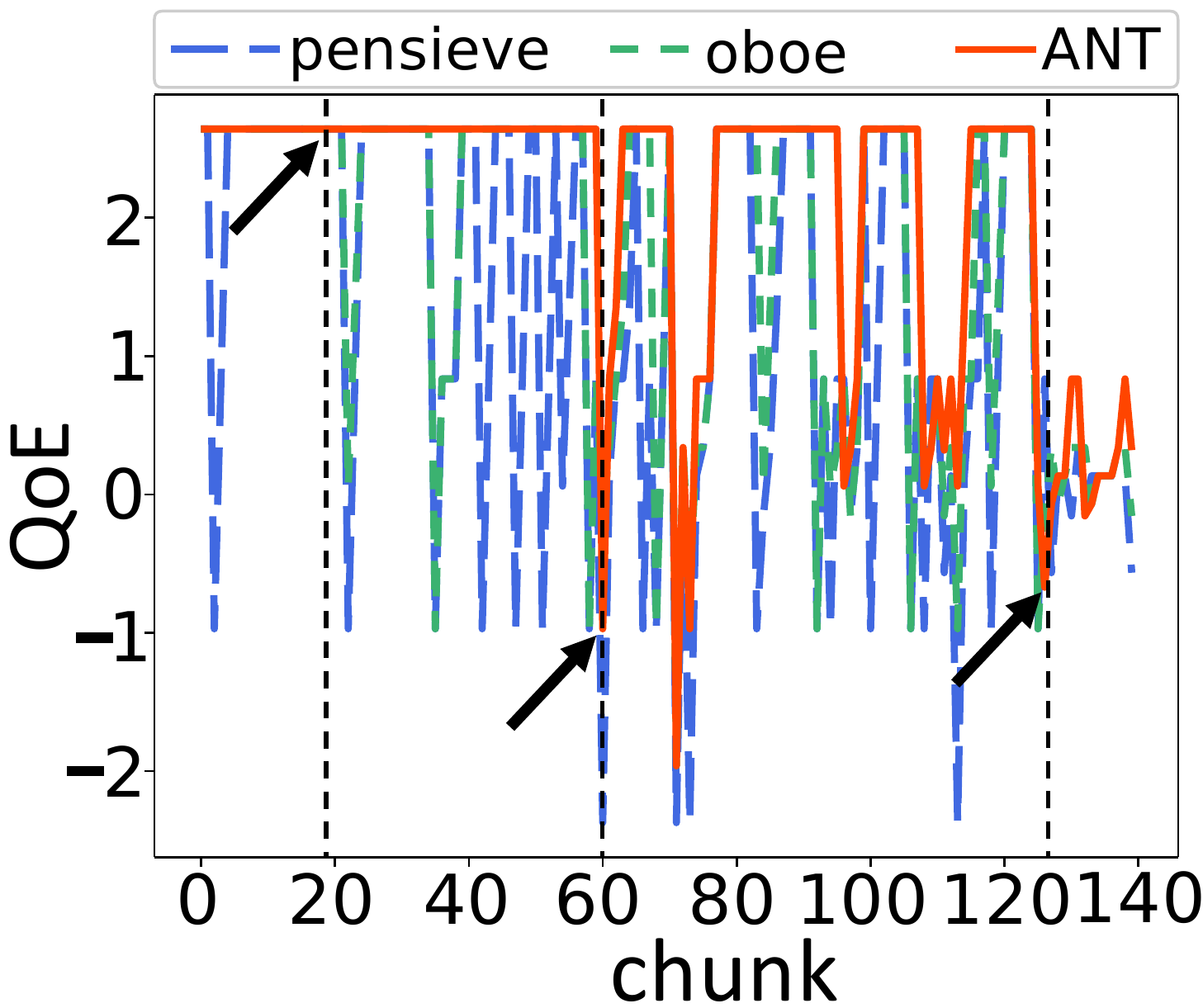} }\hspace{-1mm}\hspace{-1mm}
    \subfigure{  \includegraphics[width=0.47\linewidth]{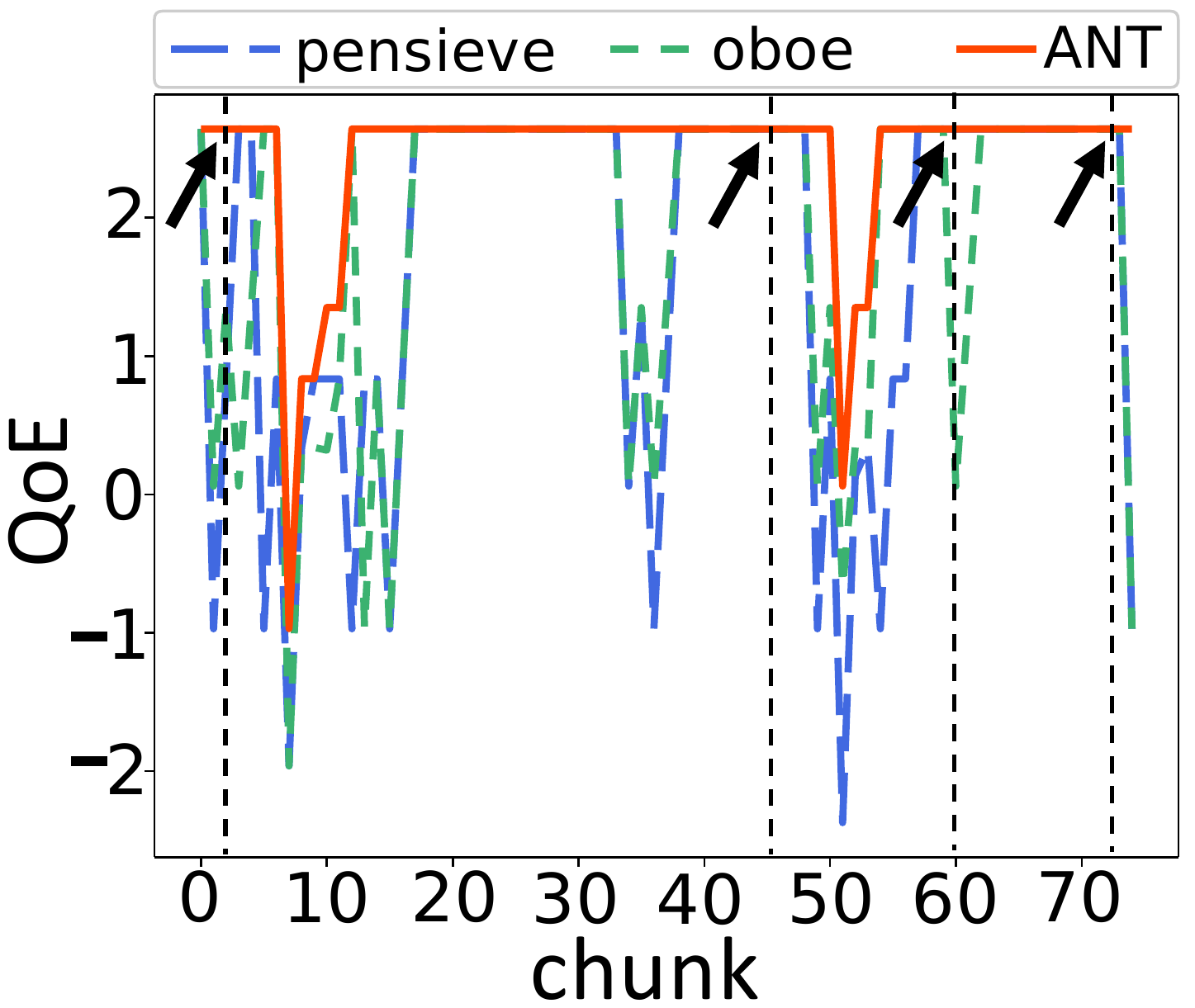}  }
    \subfigure{  \includegraphics[width=0.47\linewidth]{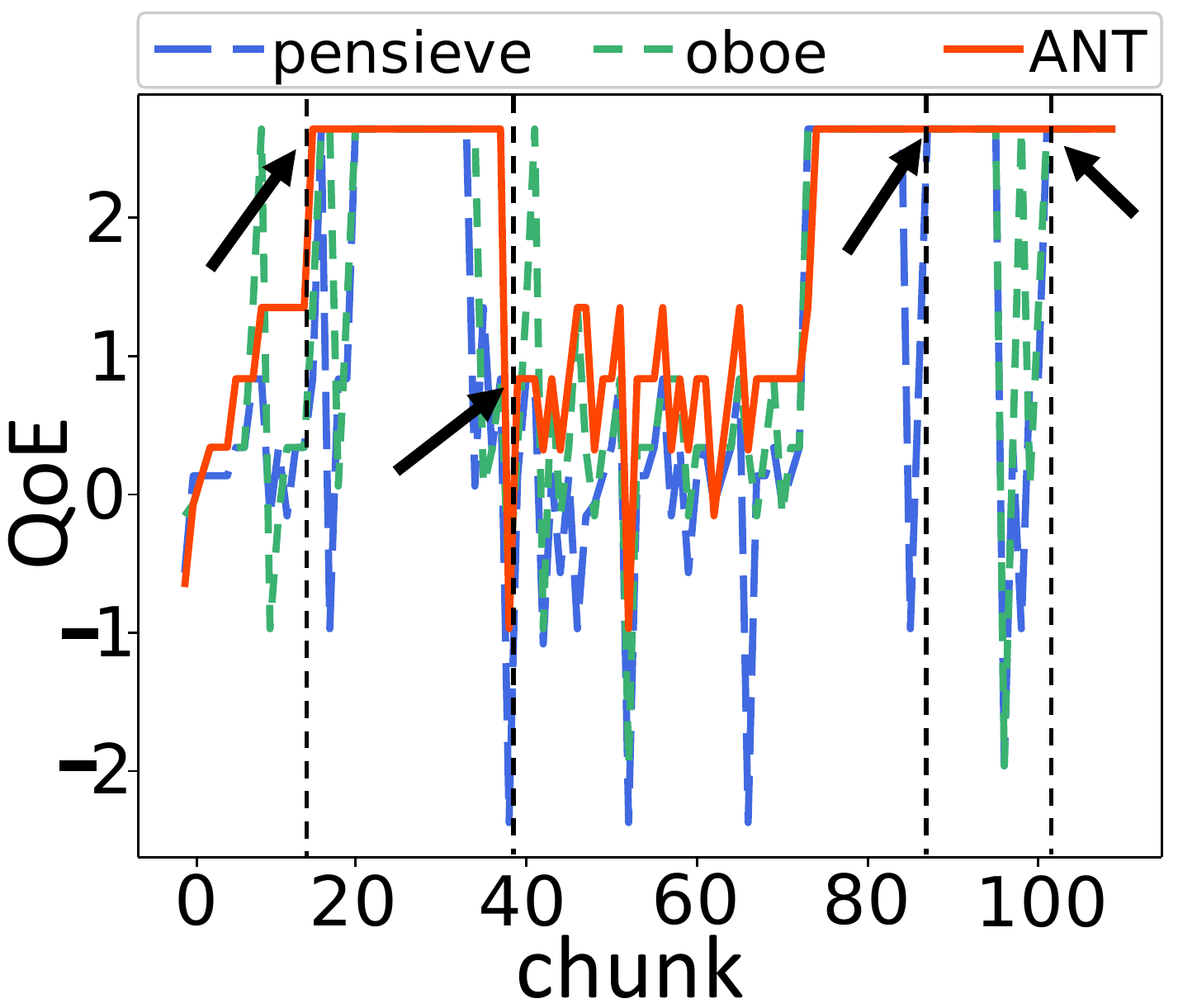} }\hspace{-1mm}
    \subfigure{  \includegraphics[width=0.47\linewidth]{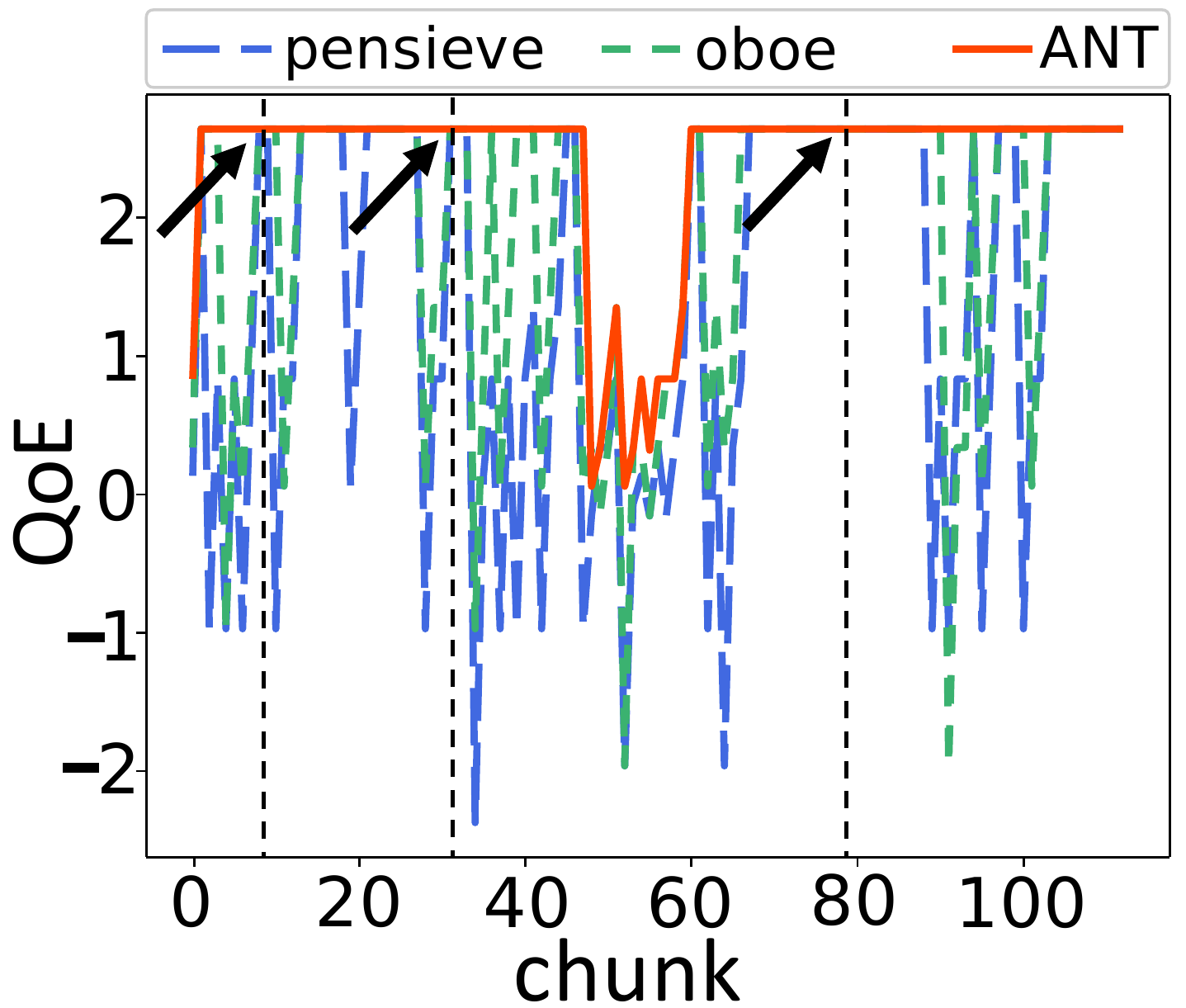}  }
    \caption{Fluctuation comparison of QoE performance.}
    \label{figure:12}
    \vspace{-5mm}
\end{figure}

Besides a good average QoE, it's necessary to maintain low instant-QoE fluctuations across adjacent video chunks. Users expect to enjoy a stable instant QoE during video streaming. {\it ANT} can acquire more stable QoE performance in the face of highly dynamic network conditions. We illustrate the stability performance of {\it ANT} and other ABR algorithms in Fig.~\ref{figure:11} and Fig.~\ref{figure:12}.

Fig.~\ref{figure:11} shows the CDF curves for the standard deviation of QoE under public traces and Tencent traces. {\it ANT} achieve a close performance of stability to Pensieve and Oboe under public traces (about 90\% traces with better stability), but outperforms them on the metric of average QoE and its individual components. For Tencent traces, {\it ANT} acquires better stability than other algorithms on almost all network conditions, and performs better on the average QoE metric and its individual components at the same time. In order to demonstrate stability directly, we also present the details of QoE fluctuations in chunk level in Fig.~\ref{figure:12}. We can find that {\it ANT} always outperforms other algorithms throughout the video chunk streaming. This is because that {\it ANT} can recognize the change of network conditions (as black arrows indicate in Fig.~\ref{figure:12}) and perform ABR model switching to maintain a high instant QoE. In contrast, Pensieve cannot tune its model parameters at all and Oboe cannot detect these network condition changes, so they face a significant performance degradation of instant QoE.

\section{Discussion}
In this paper, {\it ANT} is proposed in RL-based ABR architecture to tackle with more complex and dynamic network conditions. However, there still existing some problems to be dealt with in the future work.

\textbf{Performance on more complex network conditions.} The number of network conditions that our condition learning module can recognize is limited. Although network traces we use in this paper can cover a wide range of conditions, there is opportunity for {\it ANT} to encounter more complex network conditions in real transmission environment. In this situation, condition learning module may output inaccurate results that cause inappropriate model selection in the following bit rate decision. Moreover, new throughput data may not contain network condition labels, which brings difficulty for neural network training. Nevertheless, we devise a general ABR model and a dedicated ``uncertain'' ABR model in {\it ANT} additionally to cover above-mentioned network conditions for acceptable QoE.

\textbf{Online training for ABR models under specific network condition.} ABR server stores limited number of pre-trained ABR decision models, which correspond to considered network conditions. However,  when an absolutely different network condition appears, all pre-trained models may meet with performance degradation. At this moment, pre-trained ABR models stored in ABR server need to be refined online under this unseen network condition to learn a well-matched ABR model for bit rate decision. This requires additional studies and will be deferred as our future work.

\section{Related Work}

\textbf{ABR algorithms with fixed model.} Existing state-of-the-art ABR algorithms include rate-based (RB) algorithms~\cite{liu2011rate, jiang2012improving, sun2016cs2p, miller2016qoe, kurdoglu2016real, zou2015can}, buffer-based (BB) algorithms~\cite{huang2015buffer, huang2012confused, huang2013downton, miller2012adaptation, spiteri2016bola, beben2016abma} and hybrid control algorithms~\cite{tian2012towards, zhou2013buffer, li2014probe, wang2016squad, mansy2013sabre, yin2014toward,  yin2015control, de2013elastic, mao2017neural, akhtar2018oboe, huang2020stick, hong2019continuous, jiang2019hd3, huang2019comyco, peng2019hybrid, huang2018qarc, elgabli2020fastscan}. Due to the limited capability of rule-based algorithms, more researches focus on adopting learning-based hybrid control algorithms for ABR decision, such as Pensieve~\cite{mao2017neural} adopting A3C architecture, ~\cite{lillicrap2019continuous} with DDPG (Deep Deterministic Policy Gradient) framework, HD3 (Distributed Dueling DQN with Discrete-Continuous Hybrid Action Spaces)~\cite{jiang2019hd3} utilizing deep Q-learning algorithm~\cite{mnih2013playing}, et al. Benefiting from capabilities of neural network in feature extracting and policy learning, these algorithms tend to obtain a balance between various QoE metrics and outperform the algorithms using fixed rules. While they usually adopt single fixed model for ABR decision and do not specialize to different network conditions, {\it ANT} performs better by specializing ABR models for each network condition independently.

\textbf{Auto-tuning ABR algorithm.} In order to improve the dynamic range of ABR algorithms, Oboe~\cite{akhtar2018oboe} is proposed to auto-tune the parameters of ABR algorithms to network conditions. Oboe detects changes in network states/conditions using Bayesian change point detection algorithms based on average and standard deviation of throughput. While Oboe utilizes limited throughput statistics (i.e., average and std) with inaccurate recognition of network condition changes, {\it ANT} performs better
by learning network conditions through raw historical throughput data.

\section{Conclusion}
In this paper, we propose {~\it ANT}, an accurate network throughput learning mechanism for better adaptive video streaming. To deal with the issue of performance degradation when network condition changes frequently, a condition learning module and a condition-wised ABR decision module are devised in {~\it ANT}. By collecting and analyzing the historical throughput data, the condition learning module can accurately recognize current network condition, which is used to drive the switching of RL-based ABR models for optimal performance in the ABR decision module. Extensive experimental results verify the better performance of {\it ANT} for bit rate adaptation across a wide range of traces.

\bibliographystyle{ACM-Reference-Format}
\bibliography{acmart}

\end{document}